%% file: ssg.apj.tex
\documentclass{emulateapj-rtx4}
\usepackage{epsfig}
\usepackage{graphicx}
\usepackage{longtable}
\usepackage{natbib}
\usepackage{subfigure}
\usepackage{lscape}
\usepackage{float}

\newcommand{\Msolar}{M$_{\odot}$}
\newcommand{\Rsolar}{$R_{\odot}$}
\newcommand{\kms}{km s$^{-1}$}

\newcommand{\PRV}{$P_{\mathrm{RV}}$}
\newcommand{\PPM}{$P_{\mu}$}

\newcommand{\ClusAmp}{35.18}
\newcommand{\ClusAvgRV}{$-$47.39}
\newcommand{\ClusSig}{1.14}
\newcommand{\FldAmp}{1.55}
\newcommand{\FldAvgRV}{$-$24.66}
\newcommand{\FldSig}{46.87}

\shorttitle{WOCS. LXXI.}
\shortauthors{Milliman et al.}

\begin{document}

\title{WIYN Open Cluster Study. LXXI. Spectroscopic Membership and Orbits of NGC 6791 Sub-Subgiants}

\author{Katelyn E. Milliman\altaffilmark{1}, Emily Leiner\altaffilmark{1}, Robert D. Mathieu\altaffilmark{1}, Benjamin M. Tofflemire\altaffilmark{1}, and Imants Platais\altaffilmark{2}}
\email{milliman@astro.wisc.edu}

\altaffiltext{1}{Department of Astronomy, University of Wisconsin-Madison, 475 North Charter St, Madison, WI 53706, USA}
\altaffiltext{2}{Department of Physics and Astronomy, Johns Hopkins University, 3400 North Charles Street, Baltimore, MD 21218, USA}


\begin{abstract}
In an optical color-magnitude diagram sub-subgiants (SSGs) lie red of the main sequence and fainter than the base of the red giant branch in a region not easily populated by standard stellar-evolution pathways. In this paper, we present multi-epoch radial velocities for five SSG candidates in the old and metal-rich open cluster NGC 6791 (8 Gyr, [Fe/H] = +0.30). From these data we are able to make three-dimensional kinematic membership determinations and confirm four SSG candidates to be likely cluster members. We also identify three member SSGs as short-period binary systems and present their orbital solutions. These are the first SSGs with known three-dimensional kinematic membership, binary status, and orbital parameters since the two SSGs in M67 studied by \cite{Mathieu2003}. We also remark on the other properties of these stars including photometric variability, H$\alpha$ emission, and X-ray luminosity. The membership confirmation of these SSGs in NGC 6791 strengthens the case that SSGs are a new class of nonstandard stellar evolution products, and that a physical mechanism must be found that explains the evolutionary paths of these stars.

\end{abstract}

\section{Introduction}
Sub-subgiants (SSGs) were first discovered in the color-magnitude diagram (CMD) of the old open cluster M67 (4 Gyr; \citealt{Belloni1998, Mathieu2003}). These two stars in M67 fall to the red of the main-sequence, but well below the subgiant branch and red giant branch (RGB). This is a CMD position not easily explained given the standard theory of single-star stellar evolution. Both stars have high membership probabilities based on both proper-motion (PM) and radial-velocity (RV) data, leaving only a small probability that both are field interlopers. Notably, both of these stars are short-period binaries (2.8 days and 18 days; \citealt{Mathieu2003}), X-ray emitters \citep{vandenBerg2004}, and photometric variables \citep{vandenBerg2002}. 

Similar stars have also been discovered in globular clusters. \cite{Albrow2001} identified stars below the subgiant branch in 47 Tuc as red stragglers. In subsequent globular cluster papers stars below the subgiant branch as well as stars to the red of the RGB are referred to as either SSGs or red stragglers; see \cite{Rozyczka2012} and \cite{Cool2013} on Omega Cen; \cite{Ferraro2003} and \cite{Cohn2010} on NGC 6397, and more (e.g., \citealt{Mochejska2004, Bassa2008, Kaluzny2009, Kaluzny2010}). In addition to their CMD location, these globular cluster SSGs or red stragglers are also identified either as photometric variables or as optical counterparts to X-ray sources.

\begin{figure*}[!ht]
\begin{center} 
\subfigure{\includegraphics[width=0.475\linewidth]{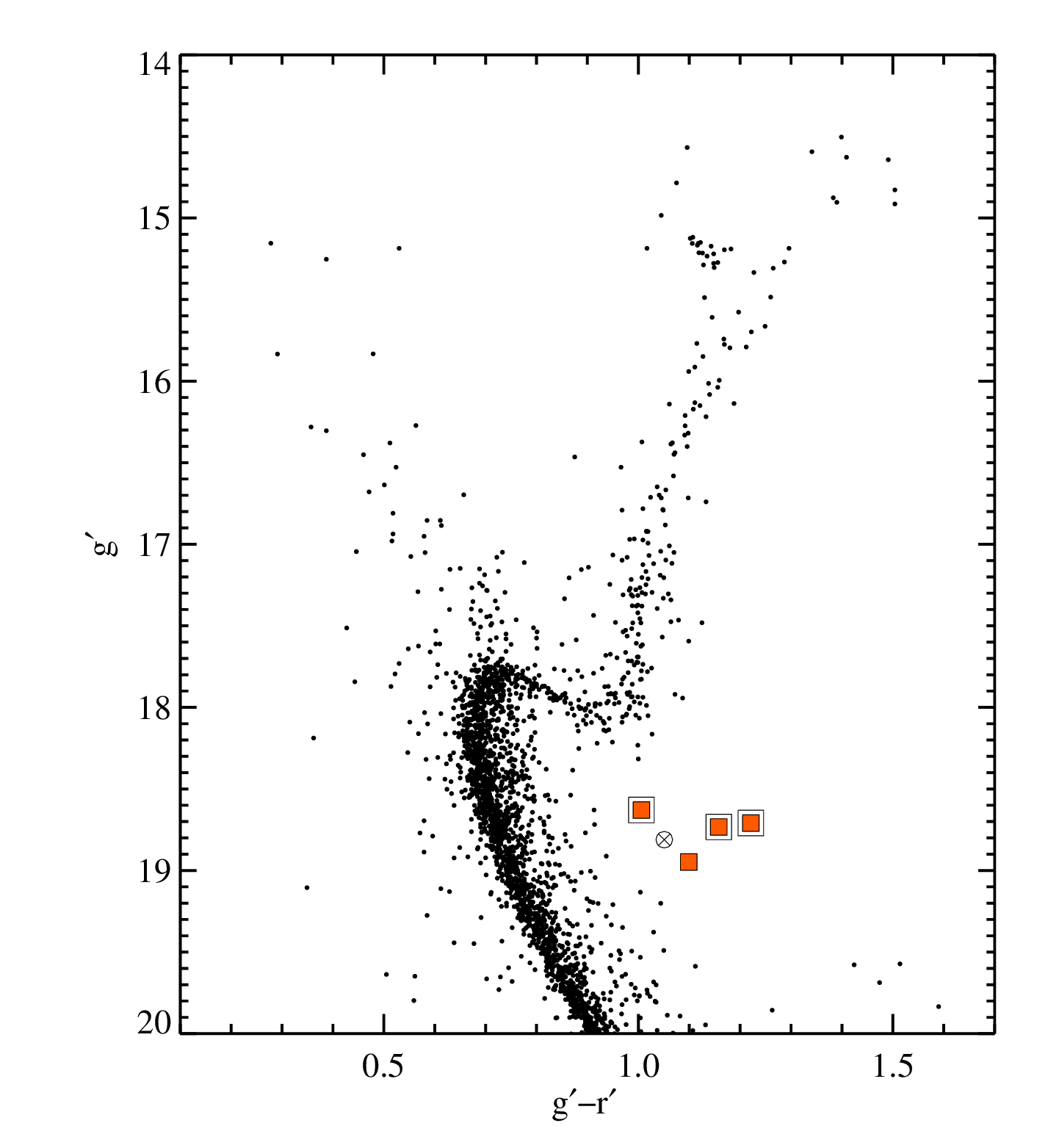}}
\subfigure{\includegraphics[width=0.475\linewidth]{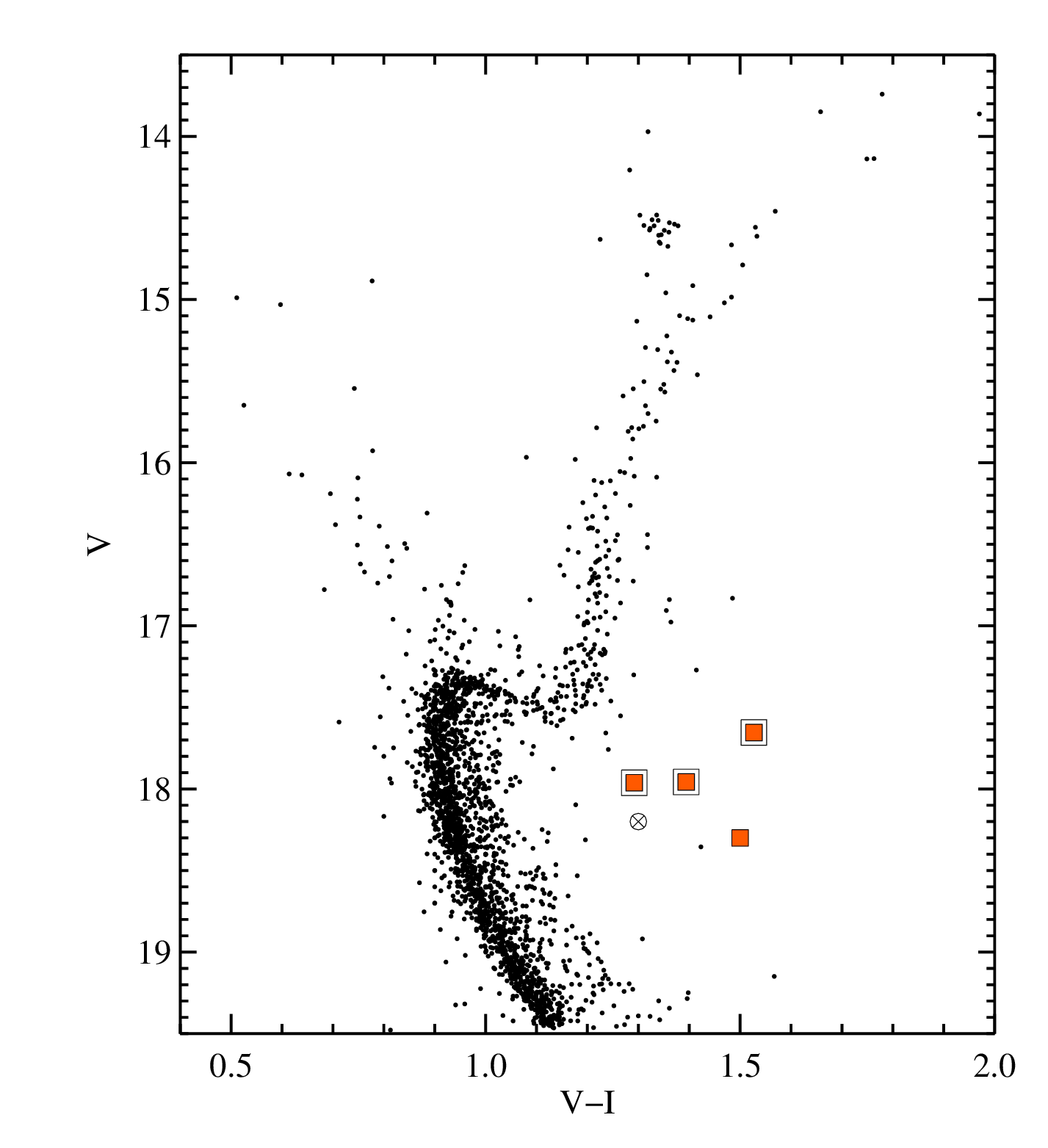}}
\caption{CMDs of the proper-motion members of NGC 6791 (\PPM~$\ge$ 95\%; \citealt{Platais2011}). The first CMD uses g$'$r$'$ from \cite{Platais2011} and the second CMD uses $VI$ photometry from \cite{Stetson2003}. Filled orange squares mark the four SSG candidates that we determine to be three-dimensional cluster members. The three with completed binary orbital solutions are enclosed by additional squares. SSG candidate WOCS 117020 that we classify as a non-member is marked by a ``$\otimes$.''}
\label{fig:cmd}
\end{center}
\end{figure*}
\begin{deluxetable*}{cccccccc}[H]
\tablewidth{\linewidth}
\tabletypesize{\scriptsize}
\tablecaption{SSG Candidate Positions and Proper Motions \label{tab:targets.pm}}
\tablehead{ \colhead{WOCS ID} & \colhead{R.A.} & \colhead{Decl.} & \colhead{$\mu_{x}$} & \colhead{$\sigma_{\mu_{x}}$} & \colhead{$\mu_{y}$} & \colhead{$\sigma_{\mu_{y}}$} & \colhead{$P_\mathrm{\mu}$} \\
\colhead{} & \colhead{(J2000)} & \colhead{(J2000)} & \colhead{(mas yr$^{-1}$)} & \colhead{(mas yr$^{-1}$)} & \colhead{(mas yr$^{-1}$)} & \colhead{(mas yr$^{-1}$)} & \colhead{(\%)}}
\startdata
   117020 & 19 21 12.38 & +37 37 49.1 & 0.41 & 0.12 & 1.03 & 0.10 & 80 \\
   130013 & 19 21 25.22 & +37 45 49.8 & 0.56 & 0.11 & 1.92 & 0.09 & 99 \\
   131020 & 19 20 10.61 & +37 51 11.2 & 0.58 & 0.27 & 1.45 & 0.15 & 96 \\
   147014 & 19 20 21.48 & +37 48 21.6 & 0.86 & 0.11 & 1.75 & 0.11 & 99 \\
   170008 & 19 20 38.88 & +37 49 04.3 & 0.87 & 0.10 & 1.53 & 0.11 & 99 
\enddata
\end{deluxetable*}

In open clusters SSGs are typically identified as being below the subgiant branch and the number of recognized SSGs is small. In addition to the two M67 SSGs, two candidate SSGs are PM members of NGC 188 \citep{Platais2003}, though their RV membership status remains uncertain \citep{Geller2008}. In NGC 6819 a possible SSG was identified in an X-ray study \citep{Gosnell2012}. Again, this star is a PM member with uncertain RV membership \citep{Platais2013, Milliman2014}. 

Broadly speaking, the populations of SSG candidates (or red stragglers) in globular and open clusters share similar characteristics. Not only do they fall in an unusual location on optical CMDs, they are also X-ray sources with $L_x \sim 10^{30}- 10^{31}$ erg s$^{-1}$ and photometric variables with periods between 1 and 15 days. Where binary status is known, they are usually found to be short-period binary systems. Most identified SSGs in open clusters are short-period binaries, suggesting that hard binaries play a role in their formation. Similarly the coronal X-rays associated with the SSGs in globular clusters may be due to rapid rotation in RS CVn-like binaries.

Beyond their association with binaries, however, the origins of sub-subgiants remain a mystery. \cite{Albrow2001} proposed that such systems may result from mass transfer, with the subsequent contraction of the primary Roche radius deflating the primary star and making it less luminous. Indeed, \cite{Mucciarelli2013} argue that the SSG companion to a millisecond pulsar in NGC 6397 has lost $\sim$70\%$-$80\% of its original mass due to mass transfer. N. Leigh (private communication) has suggested that their origin may be collisions or mergers of low-mass stars that then evolve through the sub-subgiant domain. \cite{Hurley2005} produced one sub-subgiant in their $N$-body simulation of M67 through a common-envelope event with a subgiant and main-sequence star. However, the product was not a binary. Modeling efforts have been stymied by the lack of confirmed cluster members on which to base a theory and the limited fundamental properties known for the confirmed SSG systems.

The PM study of \citet{Platais2011} identified five SSG candidates in the old, metal-rich open cluster NGC 6791 (8 Gyr, [Fe/H] = +0.3) as high-probability PM members. As part of the WIYN Open Cluster Study (WOCS; \citealt{Mathieu2000}) RV survey, we present spectroscopic observations and RV measurements of these five stars. We report each star's RV membership probability, binary status, and present the orbital solutions for three short-period binaries. We also describe the radial distribution, rotational velocity, photometric variability, H$\alpha$ profile, and X-ray emission of these SSG candidates. All this information is vital for future modeling efforts that will shed light on the origin of this new class of stars. 

\section{Sub-Subgiant Sample}
Our target list of five SSG candidates are those identified in the PM study of \cite{Platais2011}. \cite{Platais2011} report and analyze astrometry and photometry information for 58,901 objects in NGC 6791 over a circular area with a radius of 30$'$ with a faint limit of g$'$ $\sim$ 23.8. The study utilizes photographic plates from the Kitt Peak National Observatory (KPNO) 4 m and the Lick 3 m telescopes dating back to 1961, along with CCD images from the KPNO 4 m and 3.6 m Canada-France-Hawaii Telescope from 1998$-$2009. They obtain a maximum precision of 0.08 mas yr$^{-1}$, which is 1.5 \kms~at the distance of NGC 6791 (4 kpc, \citealt{Grundahl2008}). \cite{Platais2011} note an excellent separation between cluster members and field stars down to g$'$ $\sim$ 22.0.

\begin{deluxetable*}{ccccccc}
\tablewidth{\linewidth}
\tabletypesize{\scriptsize}
\tablecaption{SSG Candidate Photometry and Other Identifiers\label{tab:targets.phot}}
\tablehead{ \colhead{WOCS ID} & \colhead{g$'$} & \colhead{g$'$-r$'$} & \colhead{$V$} & \colhead{$B-V$} & \colhead{$V-I$} & \colhead{Comment\tablenotemark{a}}}
\startdata
   117020 &  18.81 & 1.05   & 18.20 & \nodata & 1.30 &  S13753 \\
   130013 &  18.71 & 1.22   & 17.65 & \nodata & 1.53 &  S15561, CX 68, 01431\_10 \\
   131020 &  18.95 & 1.10   & 18.30 & \nodata & 1.50 &  S83 \\
   147014 &  18.73 & 1.16   & 17.96 & 1.35    & 1.39 &  S746, CX 77, V59 \\
   170008 &  18.63 & 1.01   & 17.96 & 1.15    & 1.29 &  S3626, CX 30, V17
\enddata
\tablenotetext{a}{\cite{Stetson2003} IDs are preceded by an S, \cite{vandenBerg2013} IDs are preceded by CX, see \cite{deMarchi2007} for other IDs.}
\end{deluxetable*}

\cite{Platais2011} identified five stars as SSGs based on their high proper-motion membership probabilities, \PPM, and their location below the subgiant branch in their g$'$r$'$ CMD. These five SSG candidates also occupy an area away from standard stellar evolutionary tracks in a $VI$ CMD with photometry from \cite{Stetson2003} (Figure~\ref{fig:cmd}). 

For these five SSG candidates we list the coordinates and the proper motion information, including \PPM, from \cite{Platais2011} in Table~\ref{tab:targets.pm}. In Table~\ref{tab:targets.phot} we list the  g$'$r$'$ photometry from \cite{Platais2011}, the $BVI$ photometry from \cite{Stetson2003}, and other identifiers for these targets including IDs from the X-ray study of \cite{vandenBerg2013} and various photometric variability surveys.

\section{Observations, Data Reduction, and Radial Velocities}
We obtained spectra of these five SSG candidates beginning in 2014 July with the Hydra Multi-Object Spectrograph (MOS; \citealt{Barden1994}) on the WIYN\footnote{\footnotesize The WIYN Observatory is a joint facility of the University of Wisconsin-Madison, Indiana University, the National Optical Astronomy Observatory and the University of Missouri.} 3.5 m telescope. Our observations include three spectrograph setups covering three different wavelength regions and resolutions. The first observations have a spectral range of 5710$-$5950 \AA~with $R$ $\sim$ 19,000 and our second observations have a spectral range from 6450 to 6830 \AA~with $R$ $\sim$ 13,000. The third spectral range is the standard setup for the  WOCS RV survey with a wavelength range from 5000 to 5250 \AA~and $R$ $\sim$ 16,000. In Table~\ref{tab:rvs} we note the observation date and spectral setup used for each RV measurement. 

For each observation we typically obtained three science exposures totaling approximately two hours, one 100 s dome flat, and two 300 s Thorium$-$Argon emission lamp spectra (one taken before and one taken after the science exposures).

We reduced the data using standard IRAF routines. In brief, our reduction procedure was as follows. The individual frames were overscan subtracted and cosmic rays were removed from the object frames using the routine L.A. Cosmic (\citealt{Dokkum2001}). The spectra were then dispersion corrected, and the extracted spectra were flat-fielded, throughput corrected, and sky subtracted. Sky levels were determined from $\sim$15 sky fibers placed throughout the field of view that were averaged together to create a single sky spectrum. Individual exposures were co-added to create the final spectra used in our analysis. This follows the basic WOCS RV survey procedure detailed in \cite{Geller2008}, except we included a cosmic ray rejection routine and summed the individual science exposures. These steps improve the quality of these low-signal spectra and reduce the RV errors discussed in Section 4.

Using $fxcor$, we derived RVs from the centroid of a one-dimensional cross-correlation function (CCF) with an observed solar template. For the CCFs we left out regions around strong sky features (e.g.~the sodium doublet in the first spectral setup) and we avoided broad absorption lines (e.g.~H$\alpha$ in the second spectral setup). This centroid RV was converted to a heliocentric velocity and we corrected it for the unique fiber offsets of the Hydra MOS to obtain the final results listed in Table~\ref{tab:rvs}. We also include in Table~\ref{tab:rvs} the velocity error (see Section 4) and height of the CCF from $fxcor$.

\section{Radial-Velocity Error}
Various factors affect the precision of our RV measurements. Some of these factors are intrinsic to the star, including magnitude and the broadness of the spectral lines. Other factors depend also on the quality of the individual observation, including the signal-to-noise and the CCF peak height. \cite{Geller2008} includes an extensive analysis of the quality of the data and RV precision for the standard WOCS RV survey setup and targets. The usual RV precision values across all the stars and clusters observed as part of WOCS are 0.4 to 0.5 \kms~(e.g., \citealt{Geller2008, Geller2010}; \citealt{Hole2009}). 

The RVs presented in this paper have slightly worse precision. This is because the previous WOCS surveys extend to $V$ $\sim$ 16.5, while the SSG candidates in this paper range from $V$ = 17.65 to $V$ = 18.30. In addition, we have measurements from two spectral setups that have not been previously used in the WOCS survey, including one that has a lower resolution. We also do not apply the standard WOCS requirement of a CCF peak $\geq$ 0.4. Instead we include any RV measurement derived from a distinct, clear CCF peak at least 0.2 above the background noise level.

Currently, we can not repeat the $\chi^{2}$-analysis outlined in \cite{Geller2008} to quantify our RV precision because we do not have enough total observations, individual stars, or the time baseline required. However, from the analysis in \cite{Geller2008} for the standard WOCS setup, we can extrapolate that for a $V$ $\sim$ 18.0 star the RV precision is $\sim$0.7 \kms~and for a CCF peak height of $\sim$0.4 the precision is 0.9 \kms. From this we estimate our maximum precision to be approximately 0.9 \kms. 

The RV measurement errors (RV$_\mathrm{e}$) that we list in Table~\ref{tab:rvs} are the errors output by $fxcor$. These errors are computed based on the fitted peak height and width, and the asymmetric noise of the CCF. We use these errors to weight our data, as error bars in Figure~\ref{fig:ssg.plots}, and as a guide to distinguish between binary and single star as described in the next section. These errors are likely overestimates but they track the quality of the individual observations and overall they are in line with the rms residual velocities of the orbital solutions.

\input{rvs.apj.tbl}

\section{Binarity and Orbital Solutions}
\begin{figure*}[!htbp]
\subfigure{\includegraphics[width=0.33\linewidth]{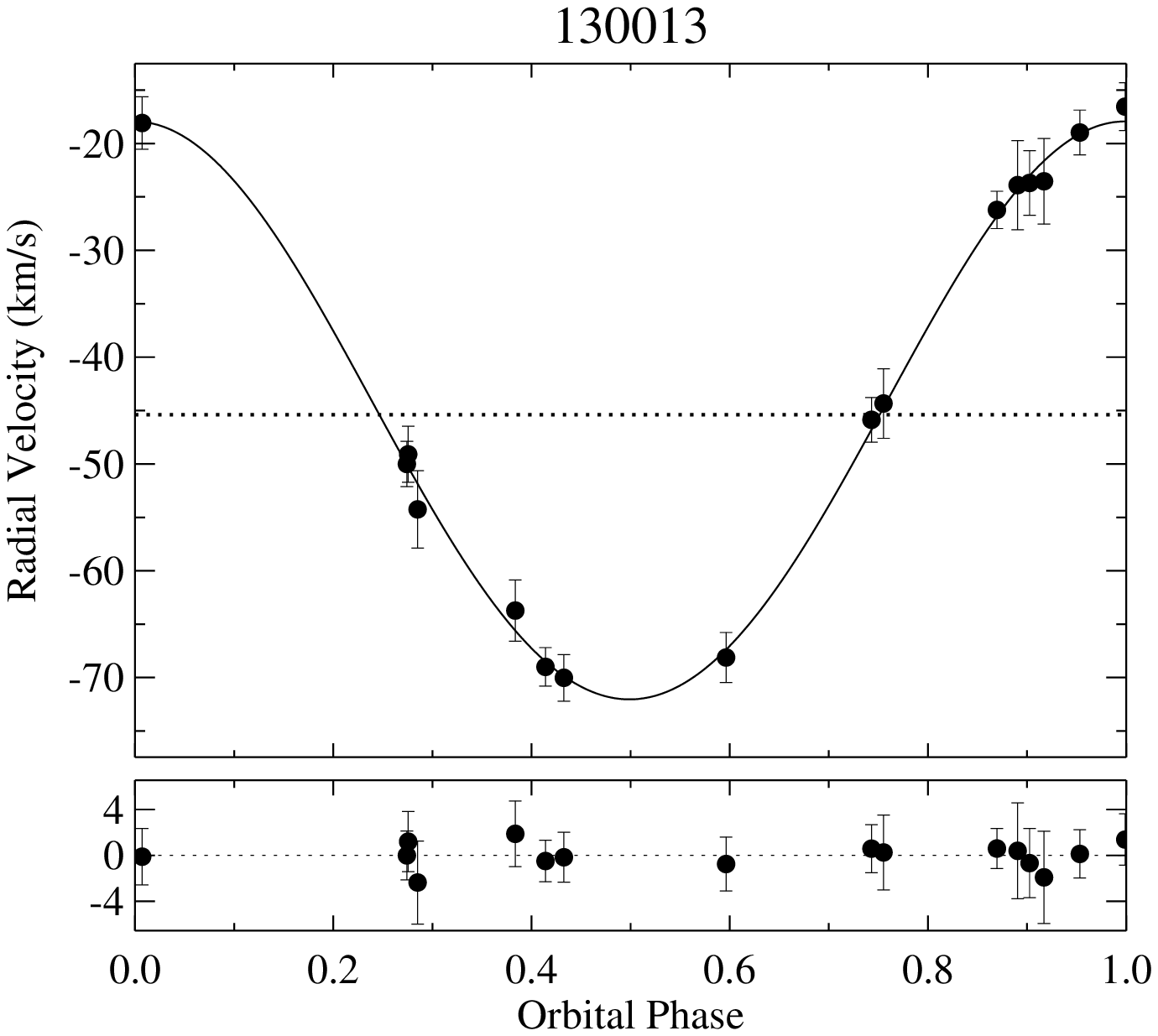}}
\subfigure{\includegraphics[width=0.33\linewidth]{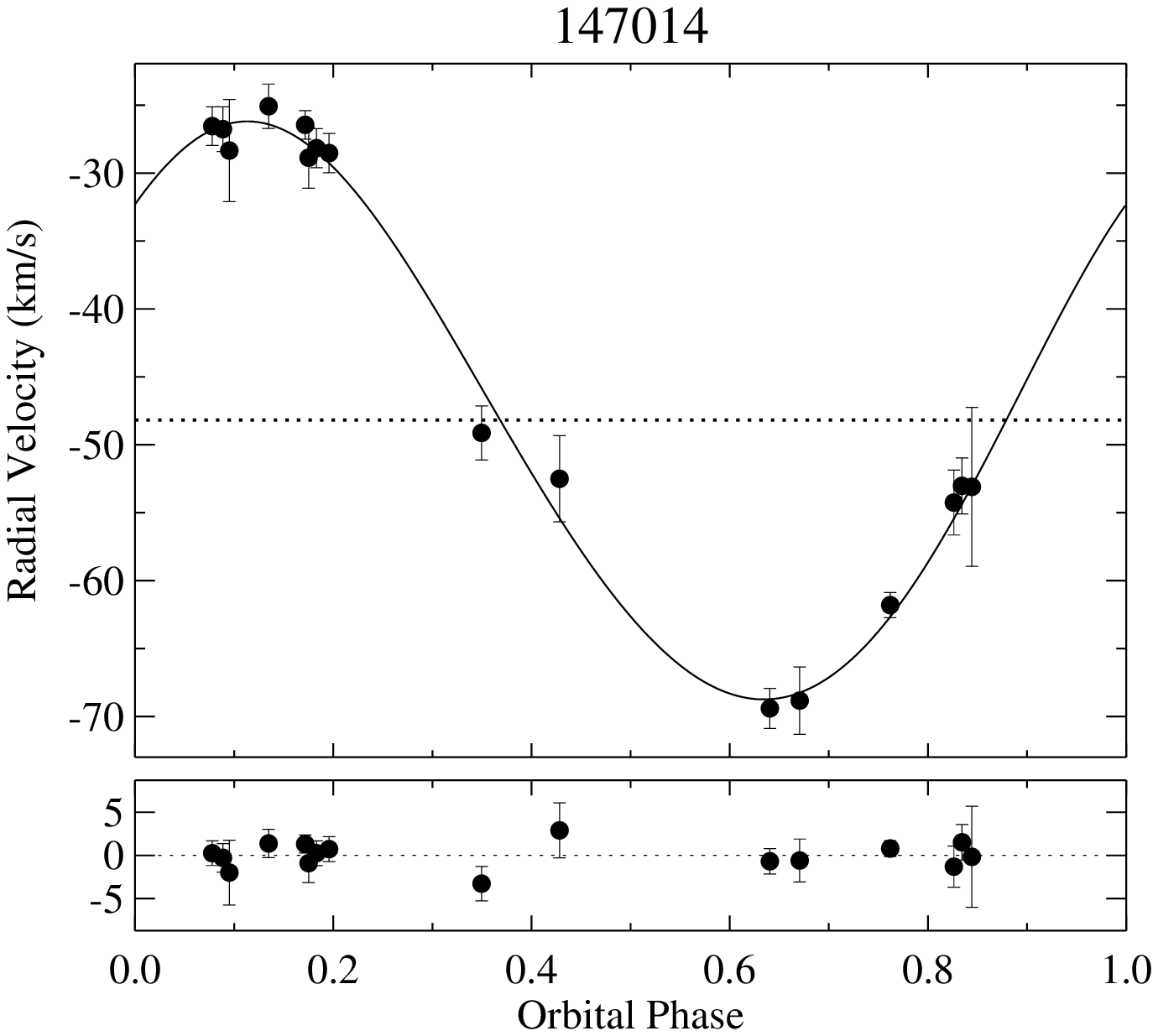}}
\subfigure{\includegraphics[width=0.33\linewidth]{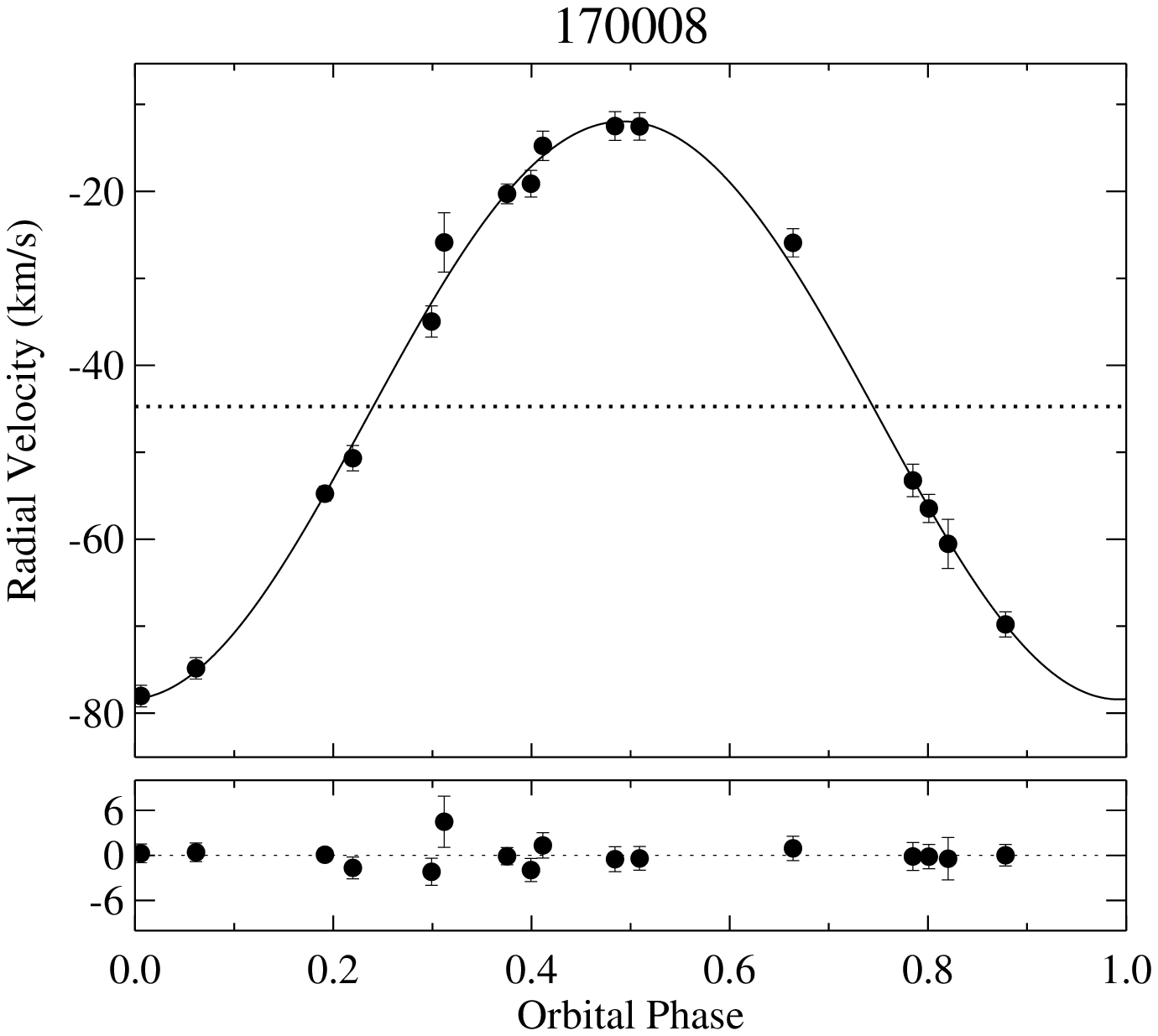}}
\caption{Orbit plots for the binary SSGs in NGC 6791. For each binary, we plot RV against orbital phase. The data points are showed with filled circles and the orbital fit to the data with the solid line. The $\gamma$-velocity is marked by the dotted line. Beneath each orbital plot, we show the O$-$C residuals from the fit. The error bars in both plots are the RV-error output by $fxcor$.}
\label{fig:ssg.plots}
\end{figure*}

\begin{deluxetable*}{cccccccccccc}
\tablewidth{\linewidth}
\centering
\tabletypesize{\scriptsize}
\tablecaption{Summary of Radial-velocity Observations\label{tab:summary}}
\tablehead{ \colhead{WOCS ID} & \colhead{$N_\mathrm{obs}$} & \colhead{$\overline\mathrm{RV}$} & \colhead{$\sigma_{\mathrm{RV}}$} & \colhead{$\overline\mathrm{RV_{e}}$} & \colhead{$\overline{v\,\mathrm{sin}\,i}$} & \colhead{$\overline{v\,\mathrm{sin}\,i}_{e}$} & \colhead{$P_\mathrm{RV}$} & \colhead{Class\tablenotemark{a}} & \colhead{$\gamma$} & \colhead{$\gamma_\mathrm{e}$} & \colhead{Comment} \\
 \colhead{} & \colhead{} & \colhead{(\kms)} & \colhead{(\kms)} & \colhead{(\kms)} & \colhead{(\kms)} & \colhead{(\kms)} & \colhead{(\%)} & \colhead{} & \colhead{(\kms)} & \colhead{(\kms)} & \colhead{}}
\startdata
117020 & 17 &    -28.0 &      1.5 &      2.1 &  $<$ 10 & \nodata &   0 &      NM & \nodata & \nodata & \nodata \\
130013 & 16 &    -43.0 &     20.1 &      2.7 &  22.4   &  1.7    &  84 &      BM &   -45.4 &     0.4 &     SB1 \\
131020 & 11 &    -49.4 &      2.3 &      2.6 &  10.2   &  0.4    &  85 &       M & \nodata & \nodata & \nodata \\
147014 & 16 &    -41.6 &     17.2 &      2.2 &  10.8   &  1.5    &  95 &      BM &   -48.2 &     0.8 &     SB1 \\
170008 & 16 &    -45.1 &     22.8 &      1.7 &  11.0   &  1.3    &  63 &      BM &   -44.7 &     0.5 &     SB1
\enddata
\tablenotetext{a}{NM: non-member; BM: binary member; M: cluster member}
\end{deluxetable*}

For each SSG candidate we list the total number of RV measurements, the weighted average RV ($\overline{\textup{RV}}$), the weighted RV standard deviation ($\sigma_{\mathrm{RV}}$), and the average velocity error ($\overline\mathrm{RV_{e}}$) in Table~\ref{tab:summary}.

Two of the SSG candidates, WOCS 117020 and WOCS 131020, have RV standard deviations approximately equal to their average RV error
($\sigma_{\mathrm{RV}}$ $\simeq$ $\overline\mathrm{RV_{e}}$). Based on the low $\overline\mathrm{RV_{e}}$ for WOCS 117020, the small range in RVs, and the time baseline of over a year of observations we conclude that this is a single star. For WOCS 131020, the large range in observed RVs, 8 \kms, is explained by the high $\mathrm{RV_{e}}$ and we also classify this star as single. However, we note that a visual inspection of the images used in the PM study of \cite{Platais2011} show this star to be elongated in roughly the E-W direction and this star is potentially a visual binary with a separation of approximately 0$\arcsec$.5 $\pm$ 0$\arcsec$.2. 

The other three SSG candidates, WOCS 130013, WOCS 147014, and WOCS 170008, have RV variations much higher than can be accounted for by RV measurement error ($\sigma_{\mathrm{RV}}$ $\gg$ $\overline\mathrm{RV_{e}}$). We conclude that these three SSG candidates are single-lined spectroscopic binaries (SB1s) and we solve for orbital solutions using the RV measurements listed in Table~\ref{tab:rvs}. In Figure~\ref{fig:ssg.plots} we plot these orbital solutions (top panel) and the observed minus computed RVs or O$-$C residuals (bottom panel). We include these O$-$C residuals and the corresponding orbital phase in Table~\ref{tab:rvs}. We list the orbital parameters of these solutions in Table~\ref{tab:orbparam}. The first row contains the orbital period ($P$), the number of orbital cycles observed, the center-of-mass RV ($\gamma$), the orbital amplitude ($K$), the eccentricity ($e$), the longitude of periastron ($\omega$), a Julian Date of periastron passage ($T_{\circ}$), the projected semi-major axis ($a\,\mathrm{sin}\,i$), the mass function ($f(m)$), the rms residual velocity from the orbital solution ($\sigma$), and the number of RV measurements ($N$). The second row contains the respective errors on each of these values where appropriate. 

These orbital solutions are robust fits to the data as seen in Figure~\ref{fig:ssg.plots} and from the small errors associated with most of the orbital parameters listed in Table~\ref{tab:orbparam}. The only parameter with large errors is the longitude of periastron, $\omega$, which is expected for such near-circular orbits.  

\begin{deluxetable*}{l r c r r r r r r r c c}
\tabletypesize{\tiny}
\tablewidth{0pt}
\centering
\tablecaption{Orbital Parameters For NGC 6791 SSG Binaries\label{tab:orbparam}}
\tablehead{\colhead{WOCS ID} & \colhead{$P$} & \colhead{Orbital} & \colhead{$\gamma$} & \colhead{$K$} & \colhead{$e$} & \colhead{$\omega$} & \colhead{$T_\circ$} & \colhead{$a\,\mathrm{sin}\,i$} & \colhead{$f(m)$} & \colhead{$\sigma$} & \colhead{$N$} \\
\colhead{} & \colhead{(days)} & \colhead{Cycles} & \colhead{(\kms)} & \colhead{(\kms)} & \colhead{} & \colhead{(deg)} & \colhead{(HJD-2400000 d)} & \colhead{(10$^6$ km)} & \colhead{(\Msolar)} & \colhead{(\kms)} & \colhead{}}
\startdata
  130013 &          7.7812 &   48.9 &           -45.4 &            27.1 &           0.015 &               0 &         56989.8 &            2.89 &         1.60e-2 &  1.35 &   16 \\
         &    $\pm$ 0.0012 & \nodata &       $\pm$ 0.4 &       $\pm$ 0.6 &     $\pm$ 0.019 &        $\pm$ 90 &       $\pm$ 2.0 &      $\pm$ 0.06 &    $\pm$ 1.0e-3 & \nodata &  \nodata \\
  147014 &          11.415 &   33.3 &           -48.2 &            21.3 &            0.05 &             320 &         56995.7 &            3.34 &         1.13e-2 &  1.84 &   16 \\
         &     $\pm$ 0.007 & \nodata &       $\pm$ 0.8 &       $\pm$ 0.7 &      $\pm$ 0.04 &        $\pm$ 60 &       $\pm$ 2.0 &      $\pm$ 0.11 &    $\pm$ 1.1e-3 & \nodata &  \nodata \\
  170008 &          5.8248 &   65.3 &           -44.7 &            33.2 &           0.013 &             180 &         56997.7 &            2.66 &         2.21e-2 &  1.87 &   16 \\
         &    $\pm$ 0.0008 & \nodata &       $\pm$ 0.5 &       $\pm$ 0.8 &     $\pm$ 0.020 &        $\pm$ 100 &       $\pm$ 1.6 &      $\pm$ 0.06 &    $\pm$ 1.5e-3 & \nodata &  \nodata 
\enddata
\end{deluxetable*}

\section{Cluster Membership}
\subsection{RV Membership Probability}
To calculate the RV membership probability, \PRV, for a given star we followed the standard WOCS procedure that is based on \cite{Vasil1958} and detailed for NGC 6791 in \cite{Tofflemire2014}. Specifically we calculate the \PRV~for each SSG candidate that is listed in Table~\ref{tab:summary} using the equation:
\begin{equation}
 P_\mathrm{RV}(v)=\frac{F_\mathrm{cluster}(v)}{F_\mathrm{field}(v) + F_\mathrm{cluster}(v)},
\end{equation}
where $F_\mathrm{cluster}$ and $F_\mathrm{field}$ are separate Gaussian functions fit to the RV distributions of the cluster and field-star populations. With the continued RV observations of NGC 6791 and more stars being included in these fits, the Gaussian parameters have shifted somewhat from the numbers published in \cite{Tofflemire2014}\footnote{Increased completeness has increased the amplitude of the cluster and field Gaussians. The other parameters remain consistent within the quoted errors of \cite{Tofflemire2014}.}. The Gaussian fit parameters used in this paper including the cluster velocity, $-$47.39 \kms, are shown in Table~\ref{tab:gauss}. For each of the three SSG candidates with completed orbital solutions we use the $\gamma$-velocity in our membership calculation and for the other two stars we use $\overline\mathrm{RV}$.

\begin{figure}[htbp]
\begin{center} 
\includegraphics[scale=0.45]{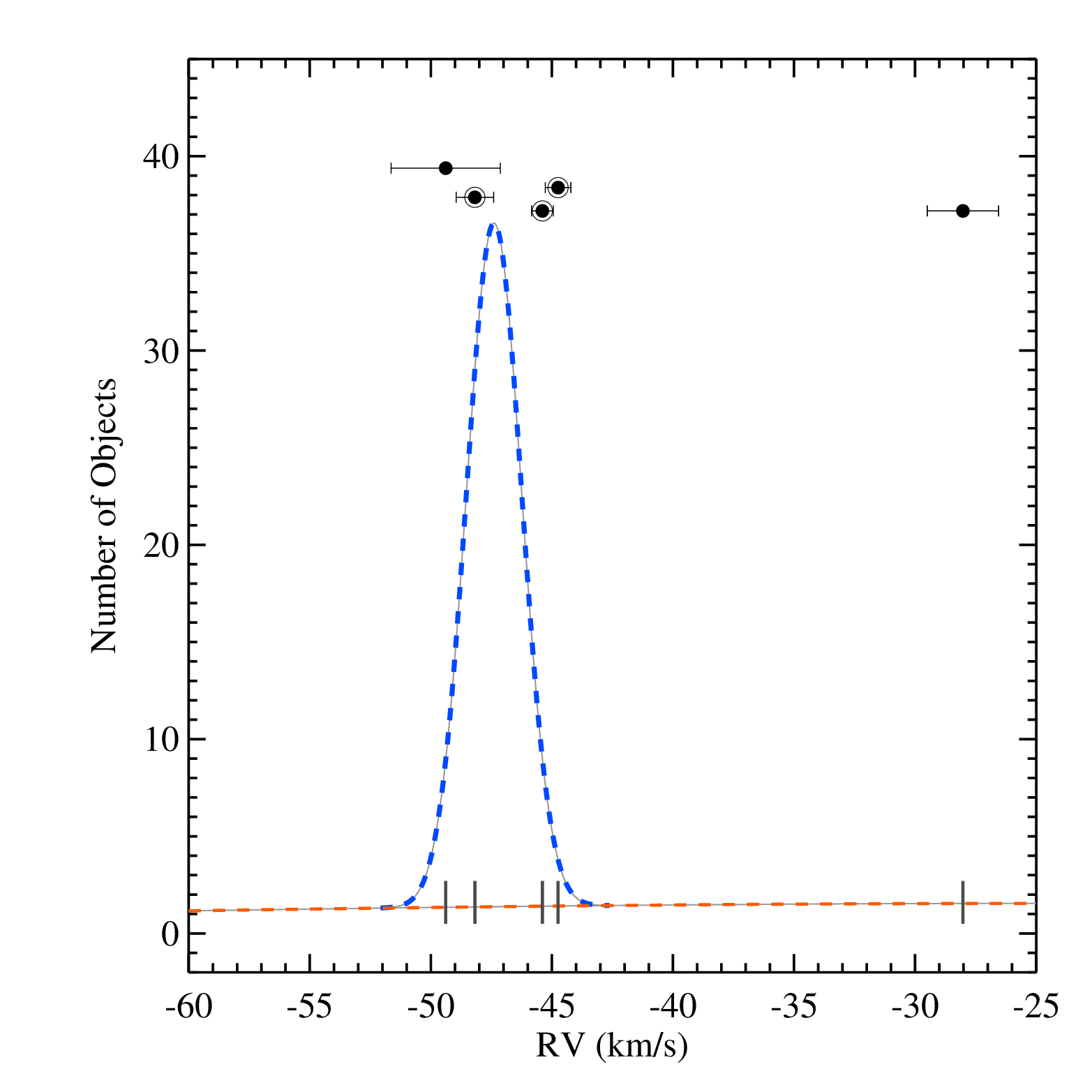}
\caption{Gaussian fits to the RV distributions of NGC 6791 (blue dashed line) and the field (orange dashed line). For the two SSG candidates without orbital solutions we overplot the average RV ($\overline{\textup{RV}}$; filled circle) and average RV-error ($\overline\mathrm{RV_{e}}$). For the three stars with completed orbital solutions we overplot the center-of-mass velocity ($\gamma$; filled and open circle) with the center-of-mass error ($\gamma_{e}$). We also mark these average or center-of-mass RVs with vertical lines near the bottom of the plot.}
\label{fig:rv.hist}
\end{center}
\end{figure}

We plot $F_\mathrm{cluster}$ and $F_\mathrm{field}$, along with the average or center-of-mass RVs of the SSG candidates in Figure~\ref{fig:rv.hist}. As seen in Figure~\ref{fig:rv.hist} and reflected in the \PRV~values, four of the SSG candidates are three-dimensional kinematic members of NGC 6791: WOCS 130013, WOCS 131020, WOCS 147014, and WOCS 170008. These four stars all have a \PPM~$\ge$ 95\% and a \PRV~above the 50\% membership threshold established by \cite{Tofflemire2014}, although WOCS 170008 with a \PRV = 63\% does lie on the tail of the probability distribution. The last star, WOCS 117020, has a \PPM~of 80\%, but the measured RVs are consistently well away from the velocity of NGC 6791 and its \PRV~is 0\%. We classify this star as a non-member. 

\subsection{Chance of Field Star Contamination}
Even with high RV and PM membership probabilities there is still a chance that any given star is actually a field contaminant. Below, we discuss the probability that the four SSGs with high \PPM~and \PRV~are field interlopers.

\begin{figure}[htbp]
\begin{center} 
\includegraphics[scale=0.5]{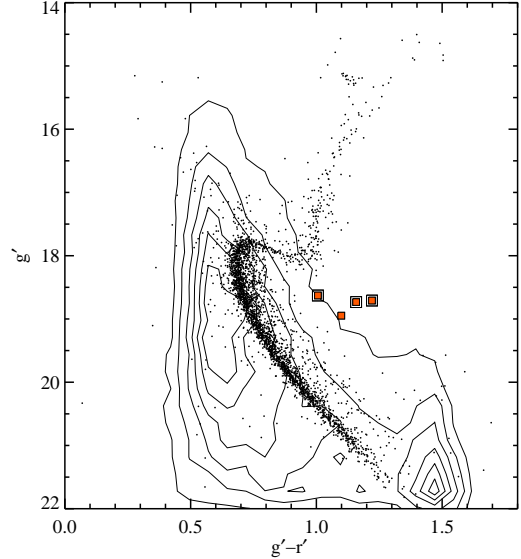}
\caption{Same as the g$'$r$'$ CMD in Figure 1. with the addition of contours indicating the field star distribution represented by \PPM~=  0\% stars. The contour levels extend from 50 to 300 in 50-star increments. The SSG stars occupy a CMD location sparsely populated by field stars, additional evidence that these stars are unlikely to be field interlopers.}
\label{fig:cmd.contours}
\end{center}
\end{figure}

First, we explore the field contamination near the SSGs in the NGC 6791 optical CMD. We use \PPM~= 0\% stars to represent the field star distribution in the direction of NGC 6791 and plot the results in Figure~\ref{fig:cmd.contours}. This shows that the SSG stars are positioned away from the two peaks in the Galactic stellar distribution and are in a color-magnitude location sparsely populated by field stars. 

Out of the 58,901 stars in the \cite{Platais2011} survey there are only 147 stars in the same area as the SSGs in the g$'$r$'$ CMD. Assuming that all 147 of these stars are not members of NGC 6791, we calculate the number expected to masquerade as three-dimensional kinematic cluster members. Integrating the PM membership probabilities of these 147 stars, we would expect to find 4.75 $\pm$ 0.50 stars to be PM members (this reflects the five SSG candidates identified by \citealt{Platais2011}).

Based on the Gaussian fit to the field around NGC 6791 described above, we expect 4.4\% of field stars to fall within the range of RVs that have \PRV~$\geq$ 50\%. Because the target selection for the \cite{Tofflemire2014} RV survey of NGC 6791 was biased toward high \PPM~stars and was limited to giants, we also calculated the RV overlap between the field and cluster based on the Gaussian fit to the field from the NGC 6819 survey of \cite{Milliman2014}. NGC 6819 is a well-populated open cluster located 4$^{\circ}$.65 away from NGC 6791. The WOCS RV survey of NGC 6819 began in 1988 and currently includes every star in the field to $\sim$2 mag below the turnoff. Incorporating the more populous and unbiased field distribution of NGC 6819, we estimate a very similar field-cluster RV overlap of 3.6\%. 

Combining the PM numbers and RV overlap statistic, of the 147 field stars with the right color and magnitude we expect 0.21 $\pm$ 0.02 would be able to masquerade as cluster members. 

Applying a Poisson distribution to this expected number we determine that the chance of detecting one field star with the cluster's PM and RV characteristics is 17\%. For detecting two stars this chance drops to 1.8\%, for three stars it is 0.13\%, and the probability we would misidentify four field stars as cluster members is 0.007\%. Based on these calculations we can confidently say that all four SSG stars with high PM and RV membership probabilities are not field contaminants and most likely three or more are indeed cluster members.

\begin{deluxetable}{crr}
\tablewidth{\linewidth}
\tablecolumns{3}
\tablecaption{Gaussian Fit Parameters for Cluster and Field RV Distributions\label{t:gauss}}
\tablehead{ \colhead{Parameter} & \colhead{Cluster} & \colhead{Field}}
\startdata
Ampl. (number) & \ClusAmp & \FldAmp \\
$\overline\mathrm{RV}$ (\kms) & \ClusAvgRV & \FldAvgRV\\
$\sigma$ (\kms) & \ClusSig& \FldSig
\enddata
\label{tab:gauss}
\end{deluxetable}

\section{Radial Distribution}
\begin{figure}
\centering
\includegraphics[width=1.1\linewidth]{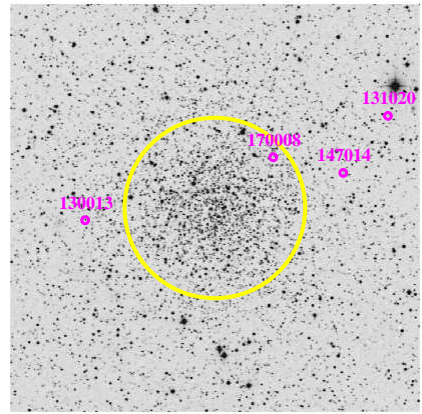}
\caption{Digital Sky Survey image of NGC 6791 overplotted with the position of the three-dimensional kinematic SSG members (magenta circles with labels) and the effective, projected ``half-mass,'' radius, $R_h$ = 4$'$.42 (\citealt{Platais2011}, yellow circle). North is up and east is to the left.}
\label{fig:ds9}
\end{figure}

The radial distribution of the four member SSGs shows clearly that they are not centrally concentrated in NGC 6791. As seen in Figure~\ref{fig:ds9}, all but one of the SSGs is outside the effective, projected ``half-mass,'' radius, $R_h$ = 4$'$.42, determined by \cite{Platais2011} from a single-mass, isotropic King model. 

\begin{figure}
\begin{center}
\includegraphics[scale=0.5]{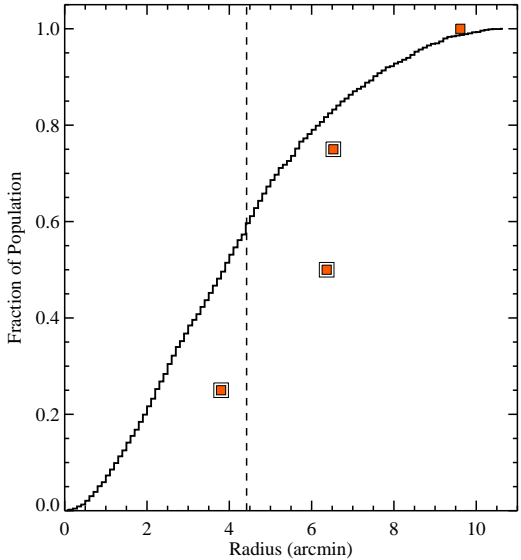}
\end{center}
\caption{Cumulative radial distribution of the high proper-motion cluster members, \PPM~$\ge$ 95\% on the main-sequence and subgiant branch, 19.5 $\ge$ g$'$ $\ge$ 17.5. The dashed line indicates the effective, projected ``half-mass,'' radius, $R_h$ = 4$'$.42  estimated by \cite{Platais2011}. The cumulative radial distribution of the four member SSGs is overplotted with orange squares and the binary systems are enclosed by an additional square.}
\label{fig:radial.dist}
\end{figure}

We compare the cumulative radial distribution of these SSG stars to the distribution of stars in NGC 6791 on the main-sequence and subgiant branch (19.5 $\ge$ g$'$ $\ge$ 17.5) with \PPM~$\ge$ 95\% (Figure~\ref{fig:radial.dist}). Approximately 60\% of these main-sequence and subgiant branch stars fall within $R_h$ = 4$'$.42, compared to only one out of the four SSGs. But a KS test between the two populations yields a $p$ = 0.11, which means we can not reject that these samples were drawn from the same distribution. Plus, with only four member stars the SSG distribution is strongly dependent on each point and there is a non-negligible chance that at least one star is a field interloper. Therefore, we are unable to draw secure conclusions from the extended radial distribution.   

\section{Photometric Variability}
Photometric variability is a general characteristic of SSGs in open and globular clusters. In this section we compile the photometric variability information available for three out of the four member SSGs in our sample.

\subsection{Observations}
NGC 6791 has been the target of many photometric studies searching for variable stars and planetary transits including \cite{Mochejska2002, Mochejska2005}, M02, M05; \cite{Kaluzny2003}, K03; \cite{Bruntt2003}, B03; and \cite{deMarchi2007}, dM07. All three binary SSGs have been classified in at least one of the above photometric studies as a rotational variable whose photometric changes are most likely the result of spot activity. \cite{deMarchi2007} do note that the light curve for WOCS 130013 is similar to that of a Cepheid variable, but because of the faintness of the target and the large distance obtained from the Period$-$Luminosity relationship ($d$ $>$ 40 Kpc) they classify it as a rotational variable. 

We list the amplitudes of variation, $\Delta$$V$, and rotation periods determined from the photometric light-curves from these studies in Table~\ref{tab:photvar.amplitude} and Table~\ref{tab:photvar.period}, respectively. We also include in Table~\ref{tab:photvar.period} our orbital periods for the three SB1's.

\subsection{Amplitude of Variation} 
WOCS 130013 has the largest $V$-band amplitude variation with $\Delta$$V$ = 0.26 found by dM07. This large amplitude is similar to the $\Delta$$V$ $\simeq$
0.2 found for the SSG S1113 in M67 (\citealt{vandenBerg2002}; \citealt{Mathieu2003}). M67's S1113 is a double-lined spectroscopic binary with a circular orbit with $P_\mathrm{orb}$ = 2.823 days (\citealt{Mathieu2003}). \cite{Mathieu2003} are able to match the large magnitude variation of S1113 with a combination of ellipsoidal variation, a large spot, and synchronous rotation. WOCS 130013 has an orbital period of 7.7815 days and we expect no significant ellipsoidal variations at that period.

WOCS 170008 has a range of $V$-band amplitude variations measured, $\Delta$$V$$_\mathrm{min}$ = 0.02 (M05) to $\Delta$$V$$_\mathrm{max}$ = 0.091 (K03). K03 suggest that both ellipsoidal effects and spot activity might be needed to explain the light curve of WOCS 170008, particularly the season-to-season shape changes and the asymmetry observed in the light
curve. But with a period of 5.8248 days we would expect ellipsoidal effects to only have a small impact on the light curve.

The measured amplitude variation for WOCS 147014 ranges from $\Delta$$V$$_\mathrm{min}$ = 0.049 (M05) to $\Delta$$V$$_\mathrm{max}$ = 0.160 (M02). For all three of these stars we suggest that the entire photometric variation is due to spot coverage and rotation. This would require large spots, especially to explain the 0.26 mag variation of WOCS 130013, perhaps indicating that these stars have substantial magnetic activity. We suggest that the different amplitudes observed are due to changes in the spot coverage between when the photometric data were taken.  

\begin{deluxetable*}{ccccccc}
\tablewidth{\linewidth}
\centering
\tabletypesize{\scriptsize}
\tablecaption{Photometric Variability: Amplitude \label{tab:photvar.amplitude}}
\tablehead{ \colhead{WOCS ID} &  \colhead{Variable ID} &  \colhead{M02} & \colhead{K03} & \colhead{B03} & \colhead{M05} & \colhead{dM07} \\
\colhead{} &  \colhead{} & \colhead{$\Delta$$V$ (mag)} & \colhead{$\Delta$$V$ (mag)} & \colhead{$\Delta$$V$ (mag)} & \colhead{$\Delta$$V$ (mag)} & \colhead{$\Delta$$V$ (mag)} }
\startdata
   130013 & 01431\_10  & \nodata  & \nodata & \nodata & \nodata & 0.26      \\
   147014 & V59        &  0.160   & \nodata & \nodata & 0.049   & \nodata   \\
   170008 & V17        &  0.032   & 0.091   & 0.036   & 0.020   & 0.04
\enddata
\end{deluxetable*}

\subsection{Rotational versus Orbital Period}\label{Prot} 
The rotation period for WOCS 130013 measured by dM07 ($P_\mathrm{rot}$ = 7.64 days) is close to our orbital period ($P_\mathrm{orb}$ = 7.7812 $\pm$ 0.0012 days). For the other variable SSGs our orbital period for each star is lower than all the measured rotational periods. For WOCS 147014 $P_\mathrm{orb}$ = 11.415 days $<$ $P_\mathrm{rot,min}$ = 13.8331 days (M05) and for WOCS 170008 $P_\mathrm{orb}$ = 5.8248 days $<$ $P_\mathrm{rot,min}$ = 6.1752 days (M02) (see Table~\ref{tab:photvar.period}). We would expect that tidal effects over the 8 Gyr age of NGC 6791 would have both circularized the binary orbit and synchronized the rotation and orbital periods. Indeed all the SSG orbits are consistent with circular systems. The discrepancy we are seeing between the orbital and rotation periods could be due to a variety of effects. 

Measurement errors could be driving the differences. The errors on our orbital period measurements are small, but none of the photometric periods have a reported error. The only statement on error comes from dM07 who note that periods over 5 days are tentative. 

Beyond measurement error, the discrepancy could be due to stellar differential rotation which, combined with high latitude spots, would result in a different measured rotation than the actual rotation period of the equator. This conclusion could be reflected in the different $P_\mathrm{rot}$ determined for WOCS 147014 and WOCS 170008 by the various photometric surveys. If the spots causing the photometric variations were at different latitudes when the various photometric observations were taken the measured $P_\mathrm{rot}$ would shift.

Or perhaps the difference between the $P_\mathrm{orb}$ and $P_\mathrm{rot}$ represents the spin-down of the primary due to radial expansion as it evolves. We urge more extensive photometric monitoring of these stars and more accurate photometric periods, as the comparison of rotational and orbital periods may provide a clue into the evolutionary histories of these SSGs.

\section{Rotational Velocities}
We derive the projected rotational velocity, $v\,\mathrm{sin}\,i$, of these stars from the FWHM of the CCF using the relationship established by \cite{Geller2010}. This relationship holds for the standard WOCS spectral setup so we include only the highest quality spectra from our most recent observations. Due to the spectral resolution of this setup we can not reliably measure $v\,$sin$\,i$ below 10 \kms. We find the weighted average projected rotational velocities to be as follows and we list them in Table~\ref{tab:summary}. WOCS 130013 $\overline{v\,\mathrm{sin}\,i}$ = 22.4 $\pm$ 1.7 \kms, WOCS 131020 $\overline{v\,\mathrm{sin}\,i}$ = 10.2 $\pm$ 0.4 \kms, WOCS 147014 $\overline{v\,\mathrm{sin}\,i}$ = 10.8 $\pm$ 1.5 \kms, and WOCS 170008 $\overline{v\,\mathrm{sin}\,i}$ = 11.0 $\pm$ 1.3 \kms~(1$\sigma$ errors).

For three of these stars the $v\,$sin$\,i$ is near our measurement lower limit of 10 \kms. WOCS 130013 stands out with its larger $v\,$sin$\,i$ = 22.4 $\pm$ 1.7 \kms. We classify this stars as a tidally synchronized SB1 ($P_\mathrm{rot}$ $\sim$ $P_\mathrm{orb}$, see Section~\ref{Prot}). We use its rotation period, $P_\mathrm{rot}$ = 7.64 days (dM07), combined with its $\overline{v\,\mathrm{sin}\,i}$ to calculate a minimum radius of $\sim$3.4$\,$\Rsolar~for this SSG. Standard stars in NGC 6791 with this radius are approximately a third of the way up the RGB based on a PARSEC isochrone (\citealt{PARSEC2012}). 

\begin{deluxetable*}{cccccccc}
\tablewidth{\linewidth}
\centering
\tabletypesize{\sciptsize}
\tablecaption{Photometric Variability: Period \label{tab:photvar.period}}
\tablehead{ \colhead{WOCS ID} &  \colhead{Variable ID} &  \colhead{M02} & \colhead{K03} & \colhead{B03} & \colhead{M05} & \colhead{dM07} & \colhead{RV Orbit Fit} \\
\colhead{} &  \colhead{} & \colhead{(days)} & \colhead{(days)} & \colhead{(days)} & \colhead{(days)} & \colhead{(days)} & \colhead{(days)} }
\startdata
   130013 & 01431\_10  & \nodata  & \nodata & \nodata & \nodata & 7.64     & 7.7812 $\pm$  0.0012  \\
   147014 & V59        & 14.4738  & \nodata & \nodata & 13.8331 & \nodata  & 11.415 $\pm$  0.007   \\
   170008 & V17        & 6.1752   & 6.20    & 6.42566 & 6.3656  & 6.523    & 5.8248 $\pm$  0.0008
\enddata
\end{deluxetable*}

\section{H$\alpha$ and X-ray Properties}

\subsection{H$\alpha$ Observations}
\begin{figure*}
\begin{center}
\subfigure{\includegraphics[width=0.45\linewidth]{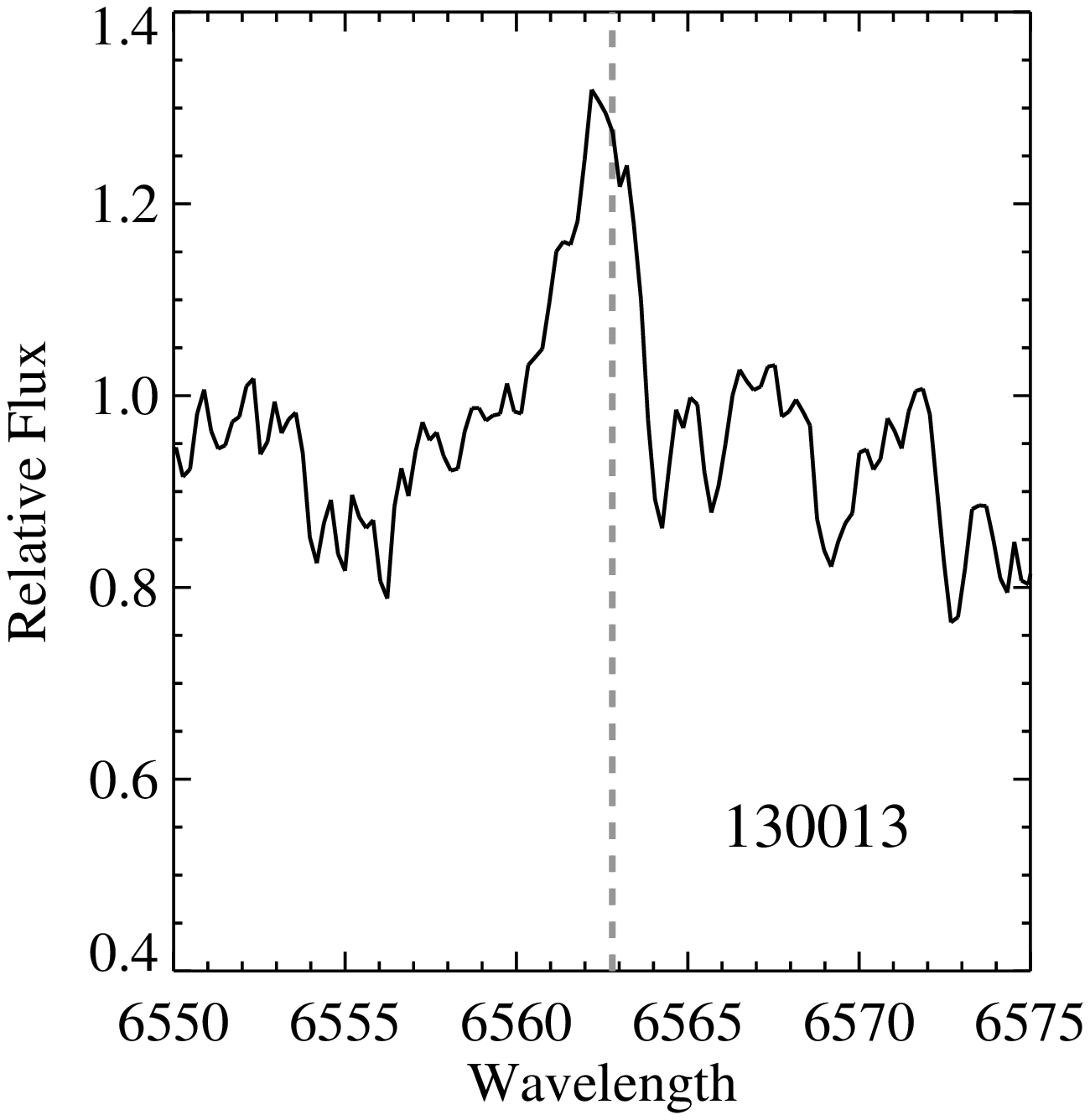}}
\subfigure{\includegraphics[width=0.45\linewidth]{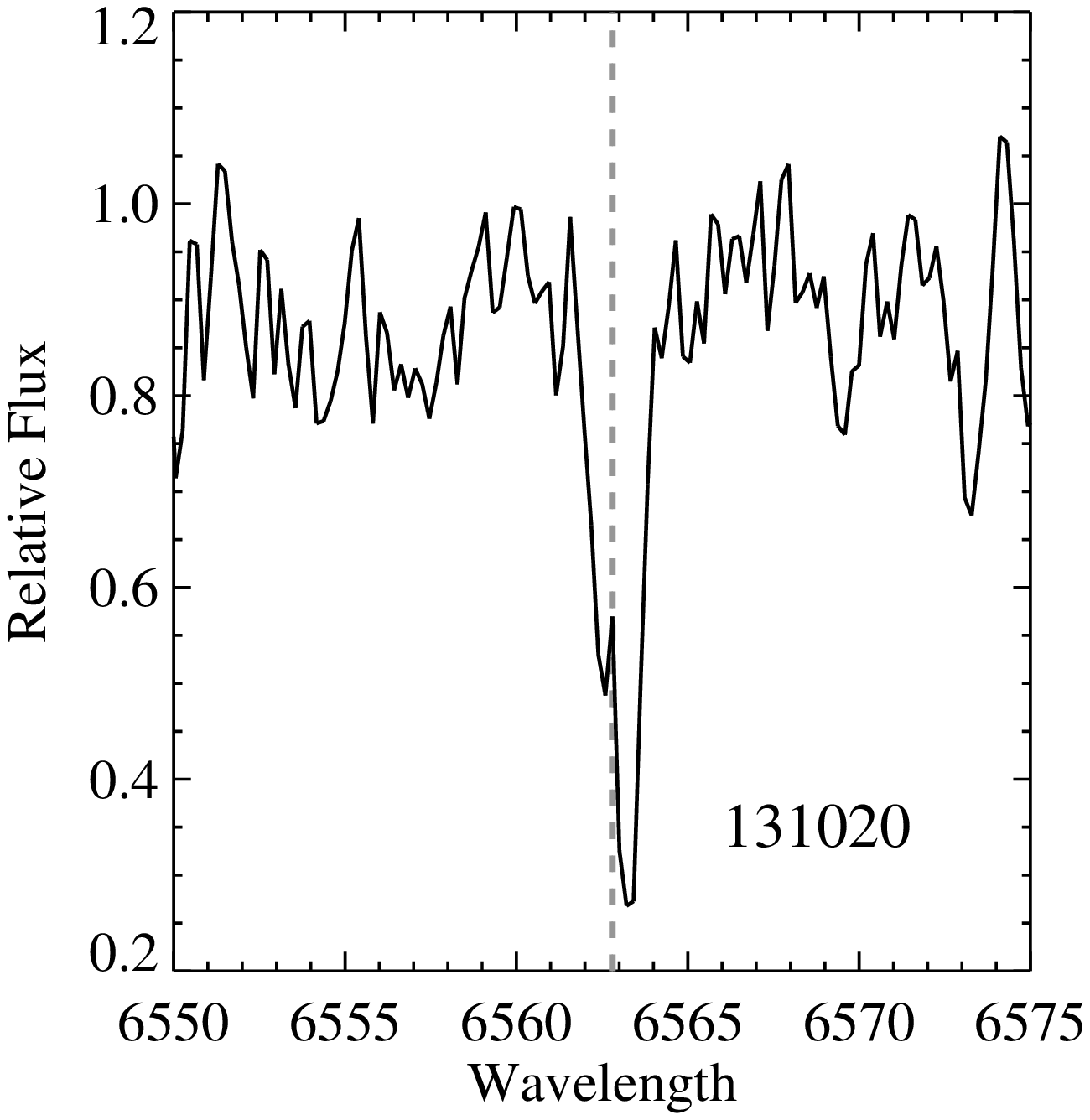}}
\subfigure{\includegraphics[width=0.45\linewidth]{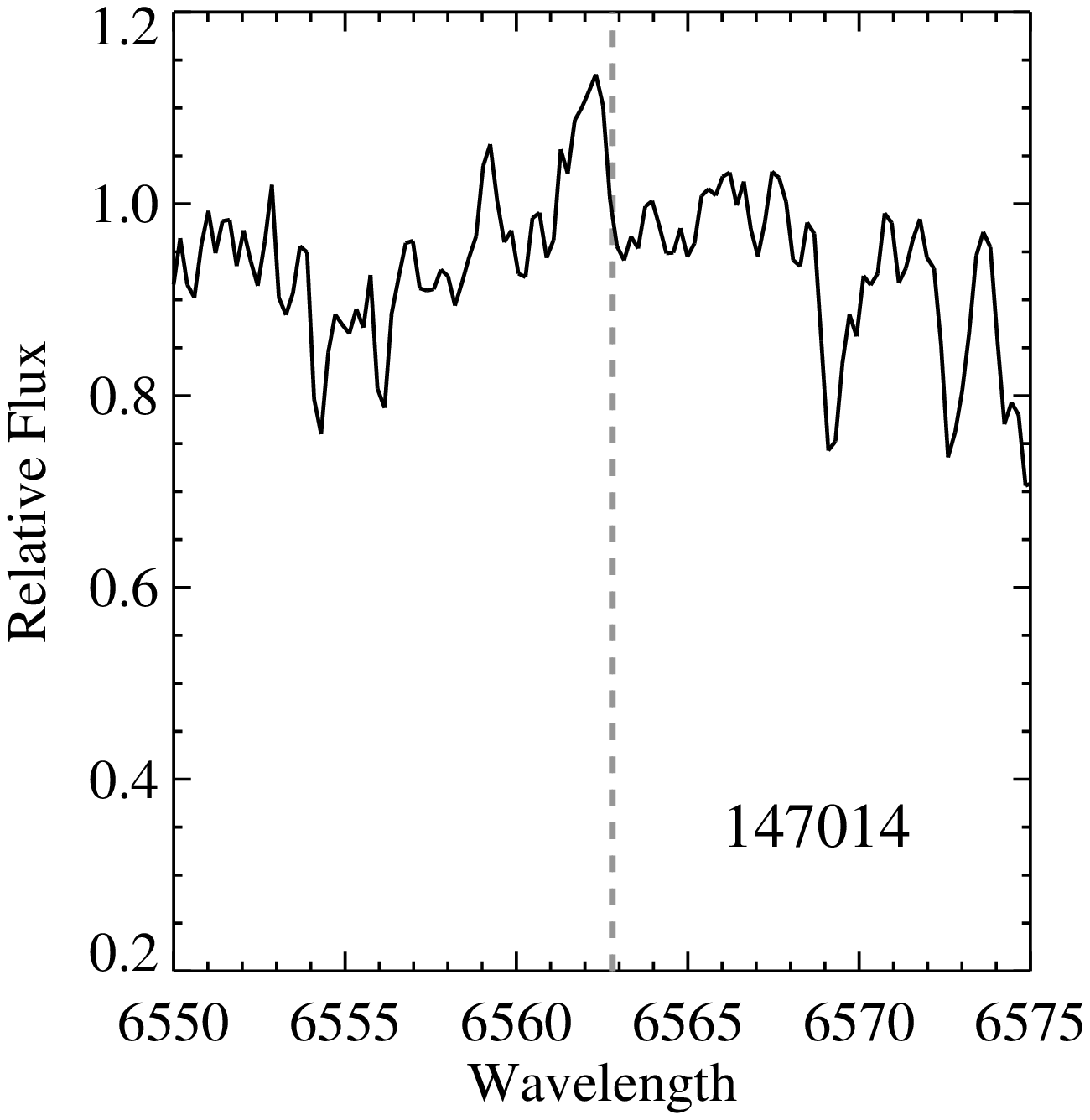}}
\subfigure{\includegraphics[width=0.45\linewidth]{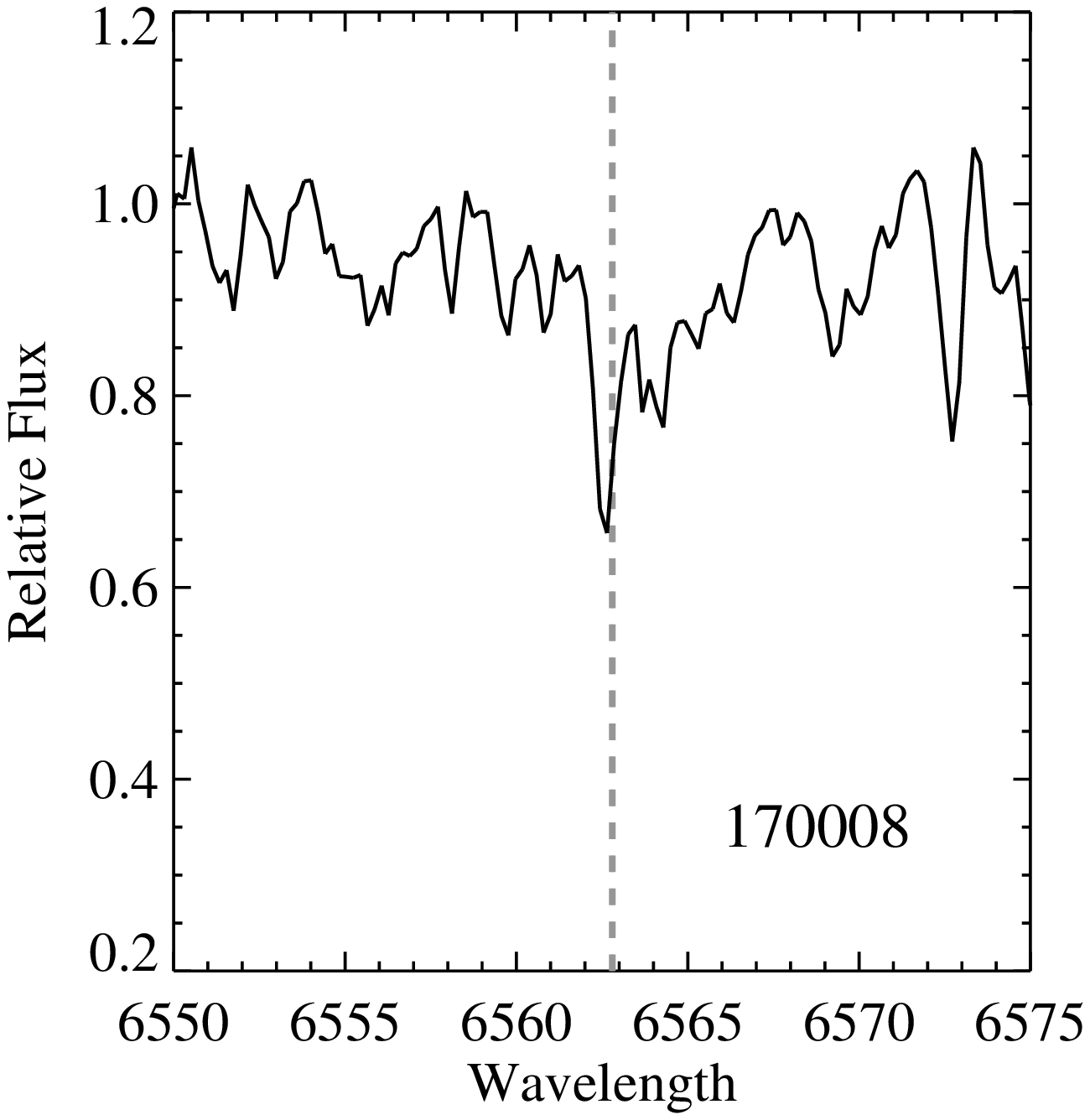}}
\end{center}
\caption{Spectra for the four member SSGs focusing on H$\alpha$. We obtained these spectra by co-adding the Doppler-shift-corrected individual spectra used for our RV measurements. The location of the H$\alpha$ feature is marked by the dashed gray line.}
\label{fig:halpha}
\end{figure*}

With the second spectral setup that extended from 6450 to 6830 \AA~we observed H$\alpha$ at a resolution of 0.515 \AA~($R$ $\sim$ 13,000). We plot the total co-added H$\alpha$ spectra of the SSG stars in Figure~\ref{fig:halpha}. To present these combined spectra we corrected for the Doppler shift of the spectra used for RV measurements (see Table~\ref{tab:rvs}) and added these RV = 0 spectra together. These co-added spectra have a total integration time of 9.4 hr and a signal-to-noise per resolution element that ranges from $\sim$27 for WOCS 131020 (g$'$ = 18.95) to $\sim$54 for WOCS 130013 (g$'$ = 18.71). The gray dashed lines in Figure~\ref{fig:halpha} mark the location of H$\alpha$ at 6562.8 \AA. 

Some of the binary SSGs in globular clusters have H$\alpha$ emission features that vary with orbital phase ($\phi$, e.g., \citealt{Sabbi2003Halpha}). We do not have observations in this spectral setup covering an entire orbit for any of the binary SSGs. The data we have extend from $\phi$ = 0.28 to 0.43 for WOCS 130013, $\phi$ = 0.09 to 0.20 for WOCS 147014, and $\phi$ = 0.30 to 0.51 for WOCS 170008. Over this range we detect no H$\alpha$ variation above the noise for any of the binary SSGs.  

WOCS 170008's spectrum has a mostly filled in H$\alpha$ absorption line when compared to the depth of $\sim$0.25 seen in WOCS 131020 and in the subgiants we observed. The spectrum of WOCS 147014 shows weak H$\alpha$ emission. WOCS 130013 has the strongest H$\alpha$ emission of the four member SSGs with an equivalent width of $\sim$0.7 \AA. This value is much lower than expected for ongoing accretion activity. This value is also much lower than the H$\alpha$ equivalent width of 15 \AA~found for the M67 SSG S1113 (\citealt{vandenBerg1999}). 

\cite{vandenBerg2013} obtained low-resolution optical spectra for three of the four SSGs in this paper. The spectra came from one of two instruments, the FAST long-slit spectrograph on the 1.5 m Tillinghast telescope on Mt. Hopkins or the fiber-fed MOS Hectospec on the 6.5 m MMT Observatory. The spectral resolution was 3 \AA~or 6 \AA, respectively. \cite{vandenBerg2013} note that WOCS 147014 has weak H$\alpha$ emission and WOCS 170008 has ``filled-in H$\alpha$?''. These descriptions match with our H$\alpha$ observations, but interestingly \cite{vandenBerg2013} make no note on the optical spectrum of WOCS 130013 which we find to have the strongest H$\alpha$ emission of the four member SSGs.

We note that the H$\alpha$ emission features in WOCS 130013 and WOCS 147014 are blueshifted. Blueshifted H$\alpha$ emission was also observed by \cite{vandenBerg1999} in the M67 SSG, S1063. We suggest this feature might be similar to the broad blueshifted H$\alpha$ component detected in RS CVn-type giants and attributed to extended prominence material or outflow structures of the chromosphere  (e.g., \citealt{Zboril2004}; \citealt{Biazzo2006}). However, the other SSG in M67, S1113, has H$\alpha$ emission centered on the velocity of the primary star and two of the SSGs in NGC 6791 do not have H$\alpha$ emission features, so blueshifted H$\alpha$ emission is not a universal property of SSGs.

\subsection{X-Ray Observations}
The X-ray study of NGC 6791 by \cite{vandenBerg2013} used the Advanced CCD Imaging Spectrometer on $Chandra$ and had a total exposure time of 48.2 ks. They detected 86 sources within 8$'$ of the cluster center including the three binary SSGs, WOCS 130013, WOCS 147014, and WOCS 170008. The non-binary member SSG, WOCS 131020, was not a detected X-ray source, but \cite{vandenBerg2013} note that it was observed with low sensitivity because of its position 9$'$ from the observations aimpoint.

\cite{vandenBerg2013} conclude from the relatively soft X-ray and the weak H$\alpha$ emission that the three detected SSGs are experiencing coronal activity and not ongoing accretion. As mentioned above, our H$\alpha$ observations agree with this conclusion. 

\section{Summary}
In this paper we perform spectroscopic follow-up on the five SSG candidates in NGC 6791 named in the PM study of \cite{Platais2011}. We find four of them to be three-dimensional kinematic members of the cluster, with high \PPM~and \PRV~values and a negligible chance of all being field interlopers.

We identify three of the four member SSGs as short-period binary systems in circular orbits. All three of these stars have been previously identified in photometric surveys as rotational variables with photometric variation likely caused by spot activity and these same three binaries were also identified as soft X-ray sources by \cite{vandenBerg2013}. The fourth member SSG, WOCS 131020, has not been identified as a photometric variable or an X-ray source. Its RV variation is approximately equal to its average RV-error and we classify it as a single star. But, we note that this star is elongated in the E-W direction on the images used in the PM study of \cite{Platais2011} and appears to be a visual binary with a separation of approximately 0$\arcsec$.5 $\pm$ 0$\arcsec$.2.

These four SSGs triple the number of SSGs in open clusters with high-likelihood three-dimensional membership probabilities, and solidify SSGs as a unique class of objects that require an evolutionary explanation. The photometric variability, soft X-ray emission, H$\alpha$ features, and binarity detailed in this paper will guide future work to narrow down likely origin theories and develop a well defined formation history for these nonstandard systems.
\\

This work was funded by the National Science Foundation grant AST-0908082 to the University of Wisconsin-Madison and several awards from the Graduate School of the University of Wisconsin-Madison.

\end{document}

%% file: rvs.apj.tbl
\begin{longtable*}{cccccccc}
\tablewidth{0pt}
\tabletypesize{\scriptsize}
\tablecaption{Radial-velocity Measurements\label{tab:rvs}}
\tablehead{ \colhead{WOCS ID} & \colhead{HJD-2,400,000} & \colhead{Spectral Setup$^{a}$} & \colhead{RV} & \colhead{$\mathrm{RV_{e}}$} & \colhead{Correlation Height} & \colhead{O$-$C} & \colhead{Phase}\\
\colhead{} & \colhead{(days)} & \colhead{} & \colhead{(\kms)} & \colhead{(\kms)} & \colhead{} & \colhead{(\kms)} & \colhead{}}
\hline \\ \multicolumn{7}{l}{{$^{a}$ Wavelength range (1) 5710$-$5950 \AA, (2) 6450$-$6830 \AA, (3) 5000$-$5250 \AA}} \\ 
\endfoot

117020   &&&&&&  \\
 & 56856.6953 &  1 &     -28.8 &      1.6 &  0.41 & \nodata & \nodata \\
 & 56856.7891 &  1 &     -29.8 &      1.2 &  0.42 & \nodata & \nodata \\
 & 56856.9023 &  1 &     -29.5 &      2.1 &  0.32 & \nodata & \nodata \\
 & 56859.6914 &  2 &     -27.7 &      2.9 &  0.42 & \nodata & \nodata \\
 & 56859.7656 &  2 &     -25.3 &      4.4 &  0.23 & \nodata & \nodata \\
 & 56860.7695 &  2 &     -27.3 &      2.1 &  0.57 & \nodata & \nodata \\
 & 56860.9141 &  2 &     -27.6 &      2.6 &  0.52 & \nodata & \nodata \\
 & 56877.7500 &  3 &     -28.4 &      1.7 &  0.47 & \nodata & \nodata \\
 & 56911.6523 &  3 &     -27.7 &      2.2 &  0.59 & \nodata & \nodata \\
 & 57088.9844 &  3 &     -25.0 &      1.8 &  0.35 & \nodata & \nodata \\
 & 57090.9766 &  3 &     -30.7 &      1.9 &  0.46 & \nodata & \nodata \\
 & 57167.7812 &  3 &     -27.3 &      1.4 &  0.60 & \nodata & \nodata \\
 & 57168.7539 &  3 &     -28.5 &      2.4 &  0.38 & \nodata & \nodata \\
 & 57168.8516 &  3 &     -28.1 &      1.5 &  0.65 & \nodata & \nodata \\
 & 57171.7812 &  3 &     -28.2 &      3.0 &  0.34 & \nodata & \nodata \\
 & 57209.8359 &  3 &     -27.2 &      1.8 &  0.61 & \nodata & \nodata \\
 & 57236.9219 &  3 &     -25.8 &      1.6 &  0.54 & \nodata & \nodata \\
130013   &&&&&&  \\
 & 56856.6953 &  1 &     -23.9 &      4.2 &  0.47 &    0.40 &    0.89 \\
 & 56856.7891 &  1 &     -23.7 &      3.0 &  0.49 &   -0.67 &    0.90 \\
 & 56856.9023 &  1 &     -23.5 &      4.0 &  0.44 &   -1.91 &    0.92 \\
 & 56859.6914 &  2 &     -49.1 &      2.6 &  0.63 &    1.20 &    0.28 \\
 & 56859.7656 &  2 &     -54.3 &      3.6 &  0.37 &   -2.36 &    0.28 \\
 & 56860.7695 &  2 &     -69.0 &      1.8 &  0.71 &   -0.49 &    0.41 \\
 & 56860.9141 &  2 &     -70.0 &      2.2 &  0.68 &   -0.16 &    0.43 \\
 & 56877.7500 &  3 &     -68.1 &      2.3 &  0.55 &   -0.75 &    0.60 \\
 & 56911.6523 &  3 &     -19.0 &      2.1 &  0.61 &    0.13 &    0.95 \\
 & 57088.9844 &  3 &     -45.9 &      2.1 &  0.49 &    0.59 &    0.74 \\
 & 57090.9766 &  3 &     -16.6 &      2.2 &  0.56 &    1.38 &    1.00 \\
 & 57167.7812 &  3 &     -26.2 &      1.7 &  0.66 &    0.60 &    0.87 \\
 & 57168.8516 &  3 &     -18.1 &      2.5 &  0.62 &   -0.11 &    0.01 \\
 & 57171.7812 &  3 &     -63.7 &      2.9 &  0.34 &    1.88 &    0.38 \\
 & 57209.8359 &  3 &     -50.0 &      2.1 &  0.61 &    0.00 &    0.27 \\
 & 57236.9219 &  3 &     -44.3 &      3.3 &  0.49 &    0.26 &    0.75 \\
131020   &&&&&&  \\
 & 56856.8086 &  1 &     -45.6 &      2.1 &  0.29 & \nodata & \nodata \\
 & 56860.7148 &  2 &     -44.4 &      3.8 &  0.36 & \nodata & \nodata \\
 & 56877.7500 &  3 &     -47.0 &      2.4 &  0.34 & \nodata & \nodata \\
 & 56911.6523 &  3 &     -47.3 &      1.4 &  0.56 & \nodata & \nodata \\
 & 57088.9844 &  3 &     -49.7 &      3.8 &  0.33 & \nodata & \nodata \\
 & 57090.9766 &  3 &     -52.4 &      2.7 &  0.44 & \nodata & \nodata \\
 & 57167.7812 &  3 &     -50.8 &      1.1 &  0.59 & \nodata & \nodata \\
 & 57168.8516 &  3 &     -48.2 &      3.4 &  0.58 & \nodata & \nodata \\
 & 57171.7812 &  3 &     -48.2 &      4.3 &  0.30 & \nodata & \nodata \\
 & 57209.8359 &  3 &     -50.9 &      1.7 &  0.45 & \nodata & \nodata \\
 & 57236.9219 &  3 &     -52.0 &      1.8 &  0.44 & \nodata & \nodata \\
147014   &&&&&&  \\
 & 56856.6953 &  1 &     -54.3 &      2.4 &  0.36 &   -1.30 &    0.83 \\
 & 56856.7891 &  1 &     -53.0 &      2.1 &  0.35 &    1.52 &    0.83 \\
 & 56856.9023 &  1 &     -53.1 &      5.8 &  0.25 &   -0.16 &    0.84 \\
 & 56859.6914 &  2 &     -26.8 &      1.6 &  0.63 &   -0.28 &    0.09 \\
 & 56859.7656 &  2 &     -28.3 &      3.8 &  0.33 &   -1.99 &    0.09 \\
 & 56860.7695 &  2 &     -28.2 &      1.4 &  0.73 &    0.26 &    0.18 \\
 & 56860.9141 &  2 &     -28.5 &      1.4 &  0.70 &    0.73 &    0.20 \\
 & 56877.7500 &  3 &     -68.8 &      2.5 &  0.45 &   -0.59 &    0.67 \\
 & 56911.6523 &  3 &     -69.4 &      1.5 &  0.57 &   -0.68 &    0.64 \\
 & 57088.9844 &  3 &     -28.9 &      2.2 &  0.32 &   -0.90 &    0.17 \\
 & 57090.9766 &  3 &     -49.1 &      2.0 &  0.50 &   -3.27 &    0.35 \\
 & 57167.7812 &  3 &     -26.5 &      1.4 &  0.65 &    0.26 &    0.08 \\
 & 57168.8516 &  3 &     -26.5 &      1.0 &  0.70 &    1.32 &    0.17 \\
 & 57171.7812 &  3 &     -52.5 &      3.2 &  0.25 &    2.90 &    0.43 \\
 & 57209.8359 &  3 &     -61.8 &      0.9 &  0.67 &    0.80 &    0.76 \\
 & 57236.9219 &  3 &     -25.1 &      1.6 &  0.52 &    1.38 &    0.14 \\
170008   &&&&&&  \\
 & 56856.6953 &  1 &     -53.2 &      1.9 &  0.43 &   -0.14 &    0.79 \\
 & 56856.7891 &  1 &     -56.5 &      1.6 &  0.43 &   -0.16 &    0.80 \\
 & 56856.9023 &  1 &     -60.5 &      2.8 &  0.34 &   -0.44 &    0.82 \\
 & 56859.6914 &  2 &     -35.0 &      1.8 &  0.60 &   -2.18 &    0.30 \\
 & 56859.7656 &  2 &     -25.9 &      3.4 &  0.33 &    4.48 &    0.31 \\
 & 56860.7695 &  2 &     -12.5 &      1.7 &  0.71 &   -0.50 &    0.48 \\
 & 56860.9141 &  2 &     -12.5 &      1.6 &  0.67 &   -0.40 &    0.51 \\
 & 56877.7500 &  3 &     -19.1 &      1.5 &  0.62 &   -1.95 &    0.40 \\
 & 56911.6523 &  3 &     -50.7 &      1.5 &  0.73 &   -1.66 &    0.22 \\
 & 57088.9844 &  3 &     -25.9 &      1.6 &  0.48 &    0.94 &    0.66 \\
 & 57090.9766 &  3 &     -78.0 &      1.2 &  0.55 &    0.27 &    0.01 \\
 & 57167.7812 &  3 &     -54.8 &      0.8 &  0.77 &    0.09 &    0.19 \\
 & 57168.8516 &  3 &     -20.3 &      1.1 &  0.79 &   -0.11 &    0.38 \\
 & 57171.7812 &  3 &     -69.8 &      1.4 &  0.47 &    0.02 &    0.88 \\
 & 57209.8359 &  3 &     -14.8 &      1.7 &  0.53 &    1.33 &    0.41 \\
 & 57236.9219 &  3 &     -74.8 &      1.2 &  0.69 &    0.41 &    0.06
\end{longtable*}

%% file: ssg.apj.bbl
\begin{thebibliography}{41}
\expandafter\ifx\csname natexlab\endcsname\relax\def\natexlab#1{#1}\fi

\bibitem[{{Albrow} {et~al}\mbox{.}(2001){Albrow}, {Gilliland}, {Brown},
  {et~al.}}]{Albrow2001}
{Albrow} M.~D., {Gilliland} R.~L., {Brown} T.~M., {et~al.}, 2001, \apj, 559,
  1060

\bibitem[{{Barden} {et~al}\mbox{.}(1994){Barden}, {Armandroff}, {Muller},
  {et~al.}}]{Barden1994}
{Barden} S.~C., {Armandroff} T., {Muller} G., {et~al.}, 1994, Proc. SPIE,
  2198, 87

\bibitem[{{Bassa} {et~al}\mbox{.}(2008){Bassa}, {Pooley}, {Verbunt},
  {et~al.}}]{Bassa2008}
{Bassa} C.~G., {Pooley} D., {Verbunt} F., {et~al.}, 2008, \aap, 488, 921

\bibitem[{{Belloni}, {Verbunt} \& {Mathieu}(1998){Belloni}, {Verbunt}, \&
  {Mathieu}}]{Belloni1998}
{Belloni} T., {Verbunt} F., {Mathieu} R.~D., 1998, \aap, 339, 431

\bibitem[{{Biazzo} {et~al}\mbox{.}(2006){Biazzo}, {Frasca}, {Catalano}, \&
  {Marilli}}]{Biazzo2006}
{Biazzo} K., {Frasca} A., {Catalano} S., {Marilli} E., 2006, \aap, 446, 1129

\bibitem[{{Bressan} {et~al}\mbox{.}(2012){Bressan}, {Marigo}, {Girardi},
  {et~al.}}]{PARSEC2012}
{Bressan} A., {Marigo} P., {Girardi} L., {et~al.}, 2012, \mnras, 427, 127

\bibitem[{{Bruntt} {et~al}\mbox{.}(2003){Bruntt}, {Grundahl}, {Tingley},
  {et~al.}}]{Bruntt2003}
{Bruntt} H., {Grundahl} F., {Tingley} B., {et~al.}, 2003, \aap, 410, 323

\bibitem[{{Cohn} {et~al}\mbox{.}(2010){Cohn}, {Lugger}, {Couch},
  {et~al.}}]{Cohn2010}
{Cohn} H.~N., {Lugger} P.~M., {Couch} S.~M., {et~al.}, 2010, \apj, 722, 20

\bibitem[{{Cool} {et~al}\mbox{.}(2013){Cool}, {Haggard}, {Arias},
  {et~al.}}]{Cool2013}
{Cool} A.~M., {Haggard} D., {Arias} T., {et~al.}, 2013, \apj, 763, 126

\bibitem[{{de Marchi} {et~al}\mbox{.}(2007){de Marchi}, {Poretti}, {Montalto},
  {et~al.}}]{deMarchi2007}
{de Marchi} F., {Poretti} E., {Montalto} M., {et~al.}, 2007, \aap, 471, 515

\bibitem[{{Ferraro} {et~al}\mbox{.}(2003){Ferraro}, {Sabbi}, {Gratton},
  {et~al.}}]{Ferraro2003}
{Ferraro} F.~R., {Sabbi} E., {Gratton} R., {et~al.}, 2003, \apjl, 584, L13

\bibitem[{{Geller} {et~al}\mbox{.}(2010){Geller}, {Mathieu}, {Braden},
  {et~al.}}]{Geller2010}
{Geller} A.~M., {Mathieu} R.~D., {Braden} E.~K., {et~al.}, 2010, \aj, 139, 1383

\bibitem[{{Geller} {et~al}\mbox{.}(2008){Geller}, {Mathieu}, {Harris}, \&
  {McClure}}]{Geller2008}
{Geller} A.~M., {Mathieu} R.~D., {Harris} H.~C., {McClure} R.~D., 2008, \aj,
  135, 2264

\bibitem[{{Gosnell} {et~al}\mbox{.}(2012){Gosnell}, {Pooley}, {Geller},
  {et~al.}}]{Gosnell2012}
{Gosnell} N.~M., {Pooley} D., {Geller} A.~M., {et~al.}, 2012, \apj, 745, 57

\bibitem[{{Grundahl} {et~al}\mbox{.}(2008){Grundahl}, {Clausen}, {Hardis}, \&
  {Frandsen}}]{Grundahl2008}
{Grundahl} F., {Clausen} J.~V., {Hardis} S., {Frandsen} S., 2008, \aap, 492,
  171

\bibitem[{{Hole} {et~al}\mbox{.}(2009){Hole}, {Geller}, {Mathieu},
  {et~al.}}]{Hole2009}
{Hole} K.~T., {Geller} A.~M., {Mathieu} R.~D., {et~al.}, 2009, \aj, 138, 159

\bibitem[{{Hurley} {et~al}\mbox{.}(2005){Hurley}, {Pols}, {Aarseth}, \&
  {Tout}}]{Hurley2005}
{Hurley} J.~R., {Pols} O.~R., {Aarseth} S.~J., {Tout} C.~A., 2005, \mnras, 363,
  293

\bibitem[{{Kaluzny}(2003)}]{Kaluzny2003}
{Kaluzny} J., 2003, \actaa, 53, 51

\bibitem[{{Kaluzny} \& {Thompson}(2009)}]{Kaluzny2009}
{Kaluzny} J., {Thompson} I.~B., 2009, \actaa, 59, 273

\bibitem[{{Kaluzny} {et~al}\mbox{.}(2010){Kaluzny}, {Thompson}, {Krzeminski},
  \& {Zloczewski}}]{Kaluzny2010}
{Kaluzny} J., {Thompson} I.~B., {Krzeminski} W., {Zloczewski} K., 2010, \actaa,
  60, 245

\bibitem[{{Mathieu}(2000)}]{Mathieu2000}
{Mathieu} R.~D., 2000, in ASP Conf. Ser., Vol. 198, Stellar Clusters and
  Associations: Convection, Rotation, and Dynamos, {Pallavicini} R., {Micela}
  G., {Sciortino} S., eds., p. 517

\bibitem[{{Mathieu} {et~al}\mbox{.}(2003){Mathieu}, {van den Berg}, {Torres},
  {et~al.}}]{Mathieu2003}
{Mathieu} R.~D., {van den Berg} M., {Torres} G., {et~al.}, 2003, \aj, 125, 246

\bibitem[{{Milliman} {et~al}\mbox{.}(2014){Milliman}, {Mathieu}, {Geller},
  {et~al.}}]{Milliman2014}
{Milliman} K.~E., {Mathieu} R.~D., {Geller} A.~M., {et~al.}, 2014, \aj, 148, 38

\bibitem[{{Mochejska} {et~al}\mbox{.}(2002){Mochejska}, {Stanek}, {Sasselov},
  \& {Szentgyorgyi}}]{Mochejska2002}
{Mochejska} B.~J., {Stanek} K.~Z., {Sasselov} D.~D., {Szentgyorgyi} A.~H.,
  2002, \aj, 123, 3460

\bibitem[{{Mochejska} {et~al}\mbox{.}(2004){Mochejska}, {Stanek}, {Sasselov},
  {et~al.}}]{Mochejska2004}
{Mochejska} B.~J., {Stanek} K.~Z., {Sasselov} D.~D., {et~al.}, 2004, \aj, 128,
  312

\bibitem[{{Mochejska} {et~al}\mbox{.}(2005){Mochejska}, {Stanek}, {Sasselov},
  {et~al.}}]{Mochejska2005}
{Mochejska} B.~J., {Stanek} K.~Z., {Sasselov} D.~D., {et~al.}, 2005, \aj, 129,
  2856

\bibitem[{{Mucciarelli} {et~al}\mbox{.}(2013){Mucciarelli}, {Salaris},
  {Lanzoni}, {et~al.}}]{Mucciarelli2013}
{Mucciarelli} A., {Salaris} M., {Lanzoni} B., {et~al.}, 2013, \apjl, 772, L27

\bibitem[{{Platais} {et~al}\mbox{.}(2011){Platais}, {Cudworth},
  {Kozhurina-Platais}, {et~al.}}]{Platais2011}
{Platais} I., {Cudworth} K.~M., {Kozhurina-Platais} V., {et~al.}, 2011, \apjl,
  733, L1

\bibitem[{{Platais} {et~al}\mbox{.}(2013){Platais}, {Gosnell}, {Meibom},
  {et~al.}}]{Platais2013}
{Platais} I., {Gosnell} N.~M., {Meibom} S., {et~al.}, 2013, \aj, 146, 43

\bibitem[{{Platais} {et~al}\mbox{.}(2003){Platais}, {Kozhurina-Platais},
  {Mathieu}, {et~al.}}]{Platais2003}
{Platais} I., {Kozhurina-Platais} V., {Mathieu} R.~D., {et~al.}, 2003, \aj,
  126, 2922

\bibitem[{{Rozyczka} {et~al}\mbox{.}(2012){Rozyczka}, {Kaluzny},
  {Pietrukowicz}, {et~al.}}]{Rozyczka2012}
{Rozyczka} M., {Kaluzny} J., {Pietrukowicz} P., {et~al.}, 2012, \aap, 537, A89

\bibitem[{{Sabbi} {et~al}\mbox{.}(2003){Sabbi}, {Gratton}, {Ferraro},
  {et~al.}}]{Sabbi2003Halpha}
{Sabbi} E., {Gratton} R., {Ferraro} F.~R., {et~al.}, 2003, \apjl, 589, L41

\bibitem[{{Stetson}, {Bruntt} \& {Grundahl}(2003){Stetson}, {Bruntt}, \&
  {Grundahl}}]{Stetson2003}
{Stetson} P.~B., {Bruntt} H., {Grundahl} F., 2003, \pasp, 115, 413

\bibitem[{{Tofflemire} {et~al}\mbox{.}(2014){Tofflemire}, {Gosnell}, {Mathieu},
  \& {Platais}}]{Tofflemire2014}
{Tofflemire} B.~M., {Gosnell} N.~M., {Mathieu} R.~D., {Platais} I., 2014, \aj,
  148, 61

\bibitem[{{van den Berg} {et~al}\mbox{.}(2002){van den Berg}, {Stassun},
  {Verbunt}, \& {Mathieu}}]{vandenBerg2002}
{van den Berg} M., {Stassun} K.~G., {Verbunt} F., {Mathieu} R.~D., 2002, \aap,
  382, 888

\bibitem[{{van den Berg} {et~al}\mbox{.}(2004){van den Berg}, {Tagliaferri},
  {Belloni}, \& {Verbunt}}]{vandenBerg2004}
{van den Berg} M., {Tagliaferri} G., {Belloni} T., {Verbunt} F., 2004, \aap,
  418, 509

\bibitem[{{van den Berg}, {Verbunt} \& {Mathieu}(1999){van den Berg},
  {Verbunt}, \& {Mathieu}}]{vandenBerg1999}
{van den Berg} M., {Verbunt} F., {Mathieu} R.~D., 1999, \aap, 347, 866

\bibitem[{{van den Berg} {et~al}\mbox{.}(2013){van den Berg}, {Verbunt},
  {Tagliaferri}, {et~al.}}]{vandenBerg2013}
{van den Berg} M., {Verbunt} F., {Tagliaferri} G., {et~al.}, 2013, \apj, 770,
  98

\bibitem[{{van Dokkum}(2001)}]{Dokkum2001}
{van Dokkum} P.~G., 2001, \pasp, 113, 1420

\bibitem[{{Vasilevskis}, {Klemola} \& {Preston}(1958){Vasilevskis}, {Klemola},
  \& {Preston}}]{Vasil1958}
{Vasilevskis} S., {Klemola} A., {Preston} G., 1958, \aj, 63, 387

\bibitem[{{Zboril}, {Strassmeier} \& {Avrett}(2004){Zboril}, {Strassmeier}, \&
  {Avrett}}]{Zboril2004}
{Zboril} M., {Strassmeier} K.~G., {Avrett} E.~H., 2004, \aap, 421, 295

\end{thebibliography}
